# Laser-driven solid-state synthesis of high-entropy oxides

Peng wei, Yiwen Liu, Hao Bai, Lei Zhuang[*], Hulei Yu, Yanhui Chu[*]

School of Materials Science and Engineering, South China University of Technology; Guangzhou, 510641, China.

[*]Corresponding author. lzhuang@scut.edu.cn (L.Z); chuyh@scut.edu.cn (Y.C).

**Abstract**

The vast compositional and structural landscape of high-entropy oxides (HEOs) grants them a wide range of potentially valuable physicochemical properties. However, the elemental immiscibility and crystal complexity limit their controllable synthesis. Here, we report a laser-driven solid-state synthesis technique that enables high-throughput production of HEOs with different crystal structures, including rock-salt, perovskite, spinel, fluorite, pyrochlore, tantalate, and silicate, incorporating up to 20 cationic elements. Typically, we successfully synthesize all types of high-entropy rare-earth disilicates (HEREDs), including A-, α-, β-, γ-, δ-, F-, and G-type phase structures, with up to 15 rare-earth elements in the A site and 5 transition-metal elements in the B site. Benefiting from their unique G-type phase structure and 20-cation composition, HEREDs are endowed with the new functionality of microwave absorption (effective absorption bandwidth of 4.3 GHz). Our work not only realizes the controllable synthesis of HEOs with vast compositional and structural space but also offers them new physicochemical properties, making them highly promising for a diverse array of structural and functional applications.

**Introduction**

Discovering new materials with unique physicochemical properties consistently presents vibrant opportunities for future technical advances and breakthroughs. Since their discovery in 2015, high-entropy oxides (HEOs), as a new class of inorganic materials, have attracted great attention due to their distinct compositional design philosophy and vast structural diversity.[1-3] Up to now, tremendous efforts have been made to explore HEOs with different crystal structures, including rock-salt,[1,4] perovskite,[5-7] spinel,[8] fluorite,[9,10] pyrochlore,[11] silicate,[12,13] etc. These HEOs have also been shown to possess superior properties, such as improved hardness,[9] catalytic activity,[14] ionic conductivity,[4,15] corrosion resistance,[16] and thermal stability,[17] as well as reduced polarization switching hysteresis[18] and thermal conductivity[19] compared to their single-component counterparts. These attractive properties have positioned HEOs as promising candidates in a wide range of structural and functional fields, like advanced manufacturing, aerospace, semiconductors, and energy technologies. However, it is still a significant challenge to realize the desired compositional and structural design in HEOs, which is the key to unlocking effective pathways to fine-tune their physicochemical characteristics for many applications.

To overcome the compositional and structural limitations of HEOs, considerable efforts have been dedicated to developing new synthesis techniques. Generally, the synthesis techniques can be classified into two basic categories. The first one is the moderate synthesis, including conventional solid-state reactions[1] and wet-chemistry (e.g., co-precipitation[20] and solvothermal methods[21]) that were developed in the earlier stages. These methods feature low temperatures and long synthesis timescales, commonly resulting in inevitable phase separation when confining multiple elements with severe atomic differences into a single lattice due to insufficient activation energy. To address this compositional restriction, a category of extreme synthesis involving flame aerosol[22] and Joule heat combustion approaches[23] has been exploited. Benefiting from higher temperatures that offer additional energy, it is promising to achieve a broader composition of HEOs. However, HEOs reported previously are still confined to simple high-symmetry configurations, such as rock-salt and fluorite structures. Such a structural limitation may primarily result from the contradiction that creating complex HEOs with multiple elements requires the ultrahigh temperature to provide sufficient activation energy, yet solely elevating temperatures in turn leads to severe evaporation of raw oxide materials. To this end, lasers, renowned for their high-energy density, have become an alternative way to realize ultrahigh temperatures within an extremely short timeframe. More importantly, high-energy density lasers can induce plasma plumes by breaking down surrounding gas molecules, generating a localized high-pressure condition that may help impede the evaporation of oxide raw materials.[24,25] Recently, a liquid phase-assisted laser ablation technique has been reported for synthesizing HEOs.[26] Although this technique may leverage the unique plasma plume characteristic, the compositional and structural diversity of HEOs is still restricted by the availability of precursors. To date, realizing the extensive compositional and structural landscape of HEOs remains challenging, significantly hindering the exploration of HEOs with unique physicochemical properties.

In this work, we develop a general laser-driven solid-state synthesis (LSS) technique to rapidly synthesize HEOs with vast structural and compositional space, including rock-salt, perovskite, spinel, fluorite, pyrochlore, tantalate, and silicate structures with up to 20 cationic elements. Typically, we successfully synthesize high-entropy rare-earth disilicates (HEREDs) with different complex phase types, including A-, α-, β-, γ-, δ-, F-, and G-type structures for the first time. Notably, benefiting from the unique G-type structure and the 20-cation composition (15 rare-earth (RE) cations in the A site and 5 transition-metal (TM) cations in the B site) with severe lattice distortion, HEREDs with abundant atomic vacancies are granted new functionality of microwave absorption, reaching an effective absorption bandwidth (EAB) of 4.3 GHz, with the reflection loss (RL) less than 10 dB. This work paves a new way for controllable synthesis of HEOs with vast structural and compositional space, endowing HEOs with new physicochemical properties that can facilitate further advances and breakthroughs in structural and functional fields.

**Results and discussion**

To showcase the superiority of our LSS technique over conventional synthesis methods, we first focus on RE disilicates. As shown in Fig. S1, this classic category of oxides is renowned for its diverse structural types, including low-temperature phases of A and α as well as high-temperature phases of β, γ, δ, F, and G, depending on ionic radii ($r$) of RE elements involved (see Fig. 1a).[27] Despite the appealing application potential, the synthesis of δ-, F-, and G-type high-temperature phases of HEREDs has not been reported yet, primarily due to substantial phase separation tendency arising from the considerable $r$ difference ($\sigma_r$) among multiple RE elements. To intuitively clarify the synthesis difficulties, we plotted all 1340 possible components for 4-cation equimolar HEREDs in Fig. 1b. Given that most of the reported synthesizable β and γ phases have $\sigma_r$ below this threshold (see the black cycle), a $\sigma_r$ value of 3.92% (the yellow line) is generally regarded as the threshold for phase separation in HEREDs.[13] Unfortunately, the $\sigma_r$ values for the vast majority of δ, F, and G phases exceed the 3.92% threshold (see Fig. 1b), implying that they are not synthesizable from a traditional perspective. To demonstrate the capability of LSS in overcoming these critical compositional limitations, we selected all seven phase types of HEREDs with the most pronounced $\sigma_r$ for exemplary synthesis, including β-type (Sc, Eu, Gd, Yb)$_2$Si$_2$O$_7$ (β-HERED-1), γ-type (Ce, Tm, Yb, Lu)$_2$Si$_2$O$_7$ (γ-HERED-2), δ-type (La, Tm, Yb, Lu)$_2$Si$_2$O$_7$ (δ-HERED-3), F-type (La, Ce, Yb, Lu)$_2$Si$_2$O$_7$ (F-HERED-4), G-type (La, Ce, Pr, Lu)$_2$Si$_2$O$_7$ (G-HERED-5), A-type (La, Nd, Sm, Eu)$_2$Si$_2$O$_7$ (A-HERED-6), and α-type (Eu, Gd, Dy, Ho)$_2$Si$_2$O$_7$ (α-HERED-7). The X-ray diffraction (XRD) patterns (Fig. 1c and Fig. S2) along with corresponding Rietveld refinements (low fitting parameters of $R_{wp}$ and $R_p$, see Fig. S3) suggest that the as-synthesized HERED samples are pure, with validated crystal structures and spacing groups summarized in Table S1.

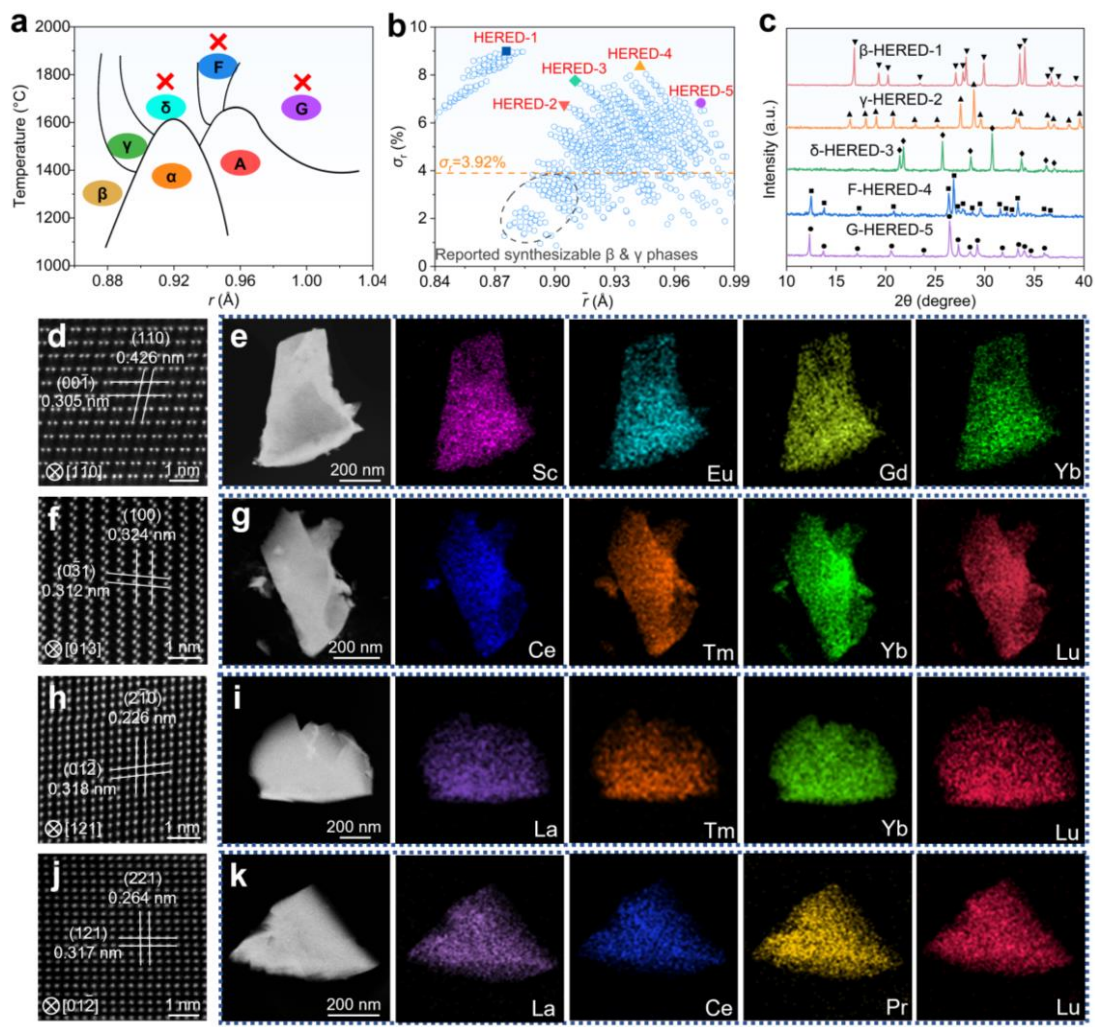

**Fig. 1 Synthesis of all seven phase types of HEREDs through the as-developed LSS.** (**a**) Polymorphic formation diagram for single-component $RE_2Si_2O_7$. (**b**) Calculated $\sigma_r$ values of all 1340 possible components in 4-cation equimolar HEREDs. (**c**) XRD patterns of as-synthesized HERED samples. Atomic-resolution HAADF-STEM images and EDS maps of the as-synthesized β-HERED-1 (**d, e**), γ-HERED-2 (**f, g**), δ-HERED-3 (**h, i**), and G-HERED-5 (**j, k**) samples.

To further verify the atomic structures of the as-synthesized HERED samples, we conducted high-angle annular dark field (HAADF) imaging using an aberration-corrected scanning transmission electron microscope (STEM). Specifically, the atomic-resolution HAADF-STEM image of the as-synthesized β-HERED-1 samples in Fig. 1d exhibits $d$-spacings of 0.305 nm and 0.426 nm, respectively, aligning well with $(00\bar{1})$ and $(110)$ planes of the β-type monoclinic structure along the $[1\bar{1}0]$ direction. The energy dispersive spectroscopy (EDS) maps in Fig. 1e indicate that Sc, Eu, Gd, and Yb elements are uniformly distributed throughout the nanoparticles in an equimolar ratio, as listed in Table S2. These results confirm the successful production of β-type HEREDs. In terms of γ-HERED-2 samples, their HAADF-STEM image in Fig. 1f displays typical $(0\bar{3}1)$ and $(100)$ planes of the γ-type monoclinic structure along the $[013]$ direction. EDS mapping results (Fig. 1g and Table S2) further demonstrate a homogenous distribution of all RE elements, validating the equimolar γ phase of HEREDs. Additionally, as shown in Fig. 1h, the *d*-spacings of δ-HERED-3 samples are measured to be 0.318 nm and 0.226 nm, matching well to $(01\bar{2})$ and $(2\bar{1}0)$ planes of the δ-type orthorhombic structure, respectively. Combined with the uniform dispersion of La, Tm, Yb, and Lu elements from the EDS maps in Fig. 1i, the δ-type phase of equimolar HEREDs can be verified. Regarding the F and G phases, it is noteworthy that they have similar structures, i.e., the F-type structure can be recognized as resulting from a slight rotation of ~1° in the *β* angle around the *c*-axis of the G-type structure[28] (see Fig. S1d and Fig. S1e). Therefore, considering the structural similarity, G-HERED-5 samples were selected as an example for further study. The HAADF-STEM image (Fig. 1j) along with EDS maps (Fig. 1k) exhibits characteristic $(121)$ and $(221)$ planes of the G-type monoclinic structure along the $[01\bar{2}]$ direction, with all RE elements homogenously distributed within the G-HERED-5 samples. These results confirm the successful synthesis of G-type HEREDs. For comparison, the synthesis of G-HERED-5 was also conducted through the conventional solid-solution reaction method. Notably, the resultant XRD patterns in Fig. S4 show that the as-synthesized samples contain impurities of X2-$RE_2SiO_5$ and γ-$RE_2Si_2O_7$. Based on these observations, it can be concluded that the synthesizable components of HEREDs have been remarkably expanded through our LSS method, effectively overcoming their compositional limitations.

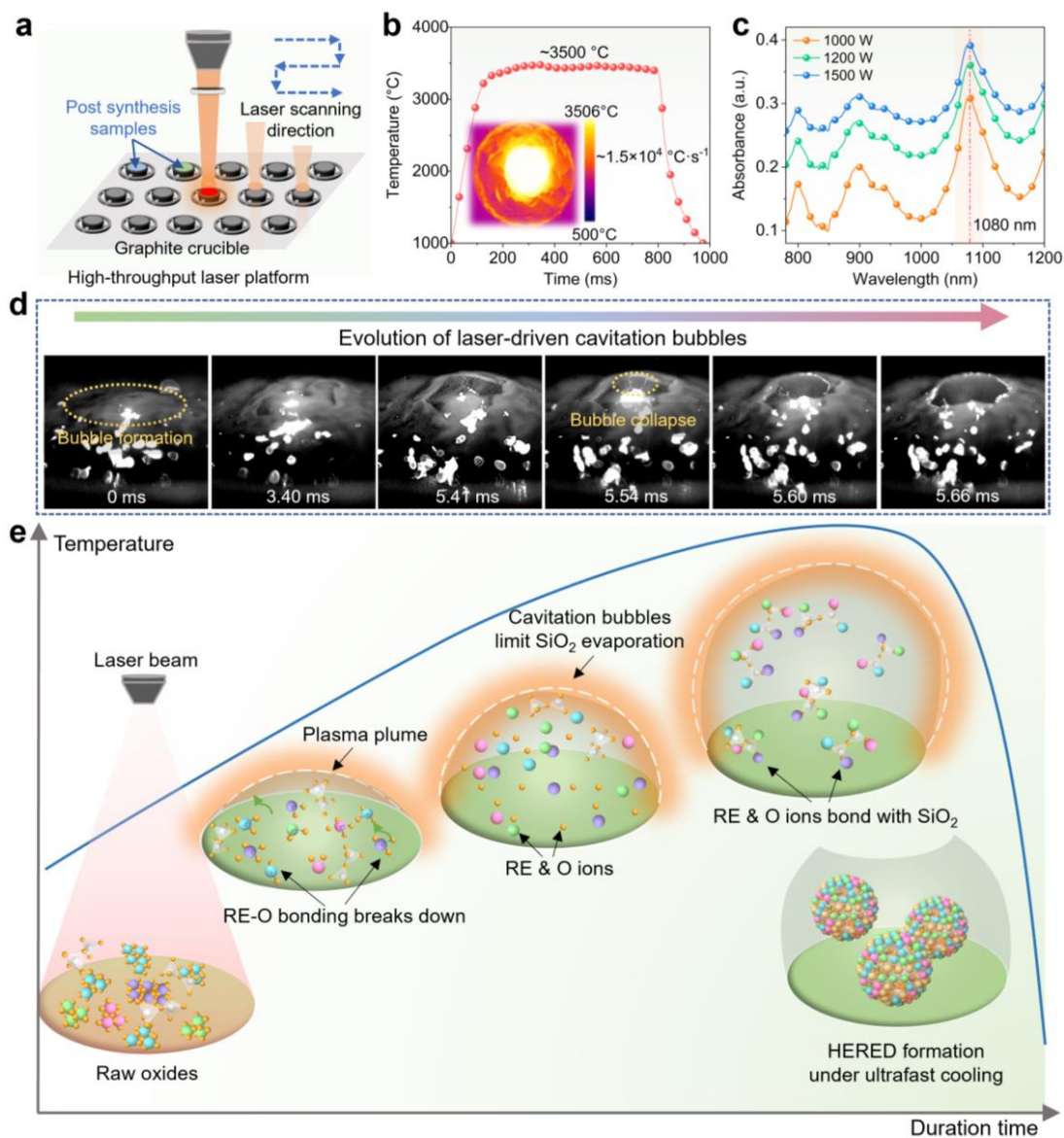

**Fig. 2 Synthesis mechanism of the as-developed LSS.** (**a**) Schematic of the working process of the self-made high-throughput LSS platform. (**b**) Temperature vs. time curve and infrared image recorded during the synthesis process. (**c**) Absorbance spectra of the as-synthesized G-HERED-5 samples. (**d**) Evolution of laser-driven cavitation bubbles captured by a high-speed camera. (**e**) Schematic of the formation process of HERED samples.

To gain insights into the unique synthesis mechanism of LSS, further studies on the synthesis process were conducted. Fig. 2a illustrates the working process of our self-made LSS platform. As the laser moves forward and scans along an "S" path, all samples can be efficiently synthesized. The temperature vs. time profile recorded during the synthesis process in Fig. 2b demonstrates an ultrafast temperature rise to ~3500 °C within milliseconds, followed by quenching with a rapid rate of ~$1.5×10^4$ °C·$s^{-1}$. The ultrahigh temperature achieved in such an extremely short time may be attributed to the strong interaction between the intense laser beam and materials, i.e., the rapid heat dissipation of massive photon energy from the laser when being absorbed and transferred into the material's lattice by carriers through electron-phonon coupling.[29] Moreover, ultraviolet-visible-near-infrared (UV-Vis-NIR) spectroscopy was utilized to validate photon absorption during the synthesis process. Fig. 2c presents the absorbance spectra of representative G-HERED-5 samples in the NIR wavelength range. Notably, a strong absorption peak can be observed at ~1080 nm, coinciding with the wavelength of the applied laser beam. Meanwhile, the peak intensity increases in proportion to the laser power tuning from 1000 W to 1500 W, providing solid evidence of photon absorption by G-HERED-5 samples. Additionally, the absorption of massive photon energy may also promote atomic vibrations, potentially leading to the excitation of atomic defects, e.g., vacancies. This speculation can be experimentally supported by the narrower bandgap of G-HERED-5 samples synthesized at 1500 W (3.85 eV) than that of samples synthesized at 1000 W (4.12 eV, see Fig. S5), primarily resulting from interband transitions caused by atomic vacancies.[30] The presence of vacancies can be further evidenced by electron paramagnetic resonance (EPR) results in Fig. S6a, which reveal that the vacancy concentrations (g=2.003, indicative of O vacancies[31]) in the as-synthesized G-HERED-5 samples increase monotonically with amplifying laser power from 1000 W to 1500 W, accompanied by the darkened colors of the as-produced samples (Fig. S6b) arising from the vacancy-enhanced light absorption.

As detailed in Fig. S7, a high-speed camera was employed to monitor the synthesis process of HEREDs through LSS. Interestingly, the real-time images (Fig. 2d) clearly captured the rapid generation and collapse of cavitation bubbles during the laser-driven synthesis process, occurring on the millisecond timescale. These bubbles may function as tiny "reaction chambers", facilitating chemical reactions between raw powders while preventing the evaporation of low-melting oxides, e.g., $SiO_2$ (~1600 °C). Such a unique synthesis characteristic distinguishes LSS from conventional synthesis methods. As illustrated in Fig. 2e, an ultrahigh temperature up to ~3500 °C can be attained instantly once the laser beam focuses on raw oxides. At such an extreme temperature, RE oxides with weak RE-O ionic bonds decompose readily,[32] resulting in the formation of RE and O ions in a liquid state along with gaseous $SiO_2$. Meanwhile, it is notable that the high-energy-density laser beam also interacts with surrounding air molecules, generating a plasma plume. A high-pressure atmospheric condition thus forms, promoting the generation of cavitation bubbles to limit the escape of $SiO_2$. Consequently, under such extreme temperature and pressure conditions, complete solid-solution reactions between various RE and O ions along with constrained gaseous $SiO_2$ easily occur, eventually yielding a series of HEREDs as designed after ultrafast cooling.

To further showcase the broad synthesis generality of the as-developed LSS, we attempted to synthesize a wide range of HEOs featuring different structures, including rock-salt, perovskite, spinel, fluorite, and pyrochlore, as shown in Fig. 3a. The XRD patterns (Fig. 3b-3d and Fig. S8) along with corresponding Rietveld refinements (Fig. S9) demonstrate the single-phase HEOs, including high-entropy rock-salt oxides, high-entropy perovskite oxides, high-entropy spinel oxides, high-entropy fluorite oxides, and high-entropy pyrochlore oxides, have been successfully synthesized within intended crystal structures (see Table S3). Besides XRD, an aberration-corrected STEM was also employed to confirm their structures. To be specific, Fig. 3e presents the atomic-resolution HAADF-STEM image of high-entropy perovskite oxide samples, revealing (002) and (201) planes with accurate $d$-spacings of the orthorhombic perovskite structure along the [010] direction. Additionally, (200) and (022) planes of the cubic spinel structure along the [0$\bar{1}$1] direction, as well as (12$\bar{2}$) and (20$\bar{4}$) planes of the cubic fluorite structure along the [201] direction, can be observed in Fig. 3f and Fig. 3g, respectively. Combining with the EDS maps displaying a uniform distribution of each constituent element in Fig. 3h-3j and Table S4, it can be deduced that the representative high-entropy perovskite oxides, high-entropy spinel oxides, and high-entropy fluorite oxides have all been successfully produced. Moreover, other HEOs, including X1-/X2-type high-entropy RE monosilicates, high-entropy RE zirconates, and high-entropy RE tantalates, were effectively synthesized, as supported by the XRD patterns and corresponding Rietveld refinements shown in Fig. S10 and Fig. S11, respectively. According to these results, it is ascertained that a diverse array of HEOs can be efficiently synthesized using our newly developed LSS technique, thereby verifying its broad generality.

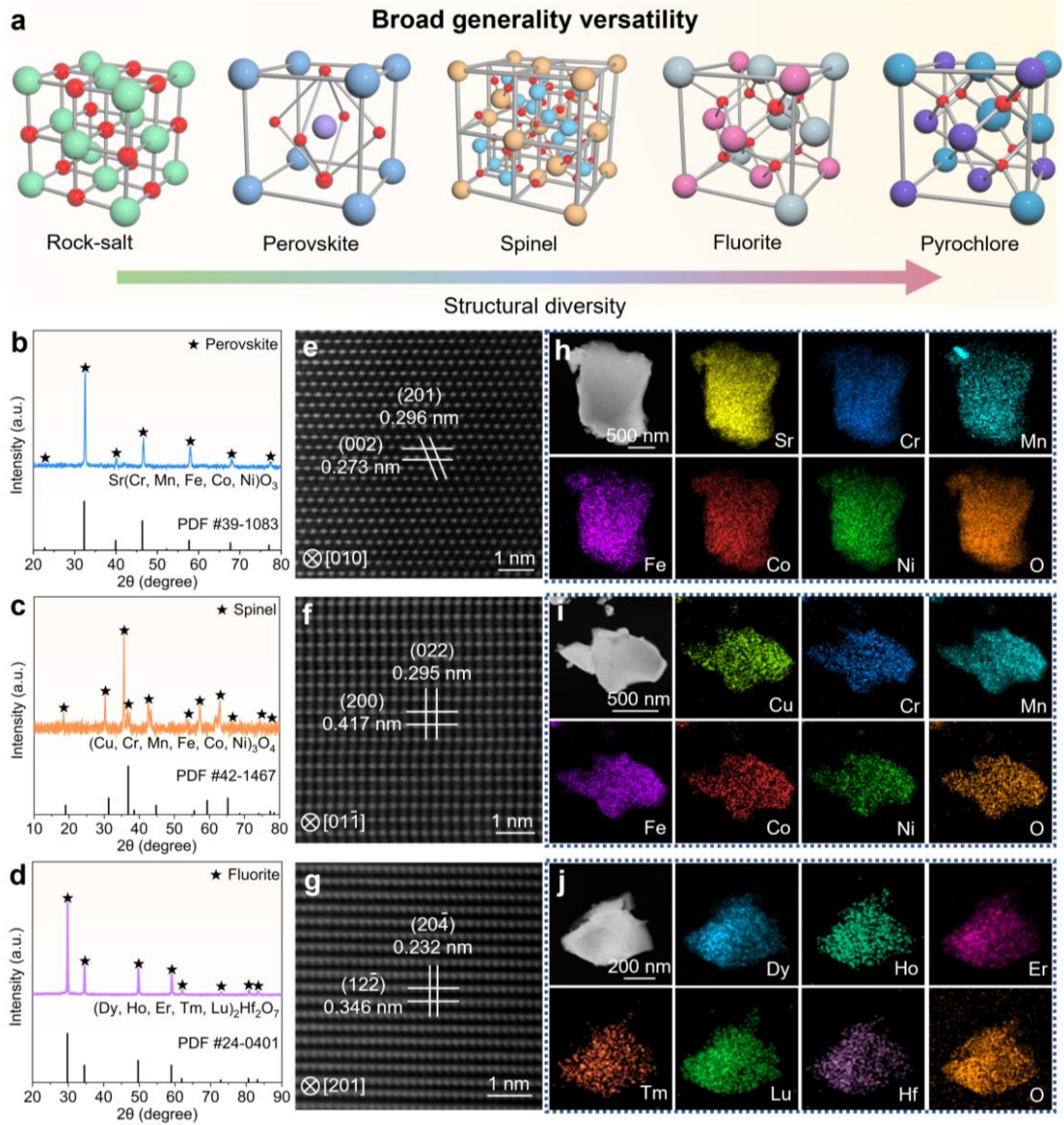

**Fig. 3 Broad synthesis generality of the as-developed LSS.** (**a**) Crystalline structures of rock-salt, perovskite, spinel, fluorite, and pyrochlore. XRD patterns (**b-d**), atomic-resolution HAADF-STEM images (**e-g**), and EDS maps (**h-j**) of the as-synthesized high-entropy perovskite oxides, high-entropy spinel oxides, and high-entropy fluorite oxides.

HEOs with more constituent elements are essential for broadening their compositional space and achieving unprecedented properties. However, realizing a uniform mixture of numerous elements in a single-phase system remains difficult, primarily due to considerable differences between elements, such as $\sigma_r$. For example, as shown in Fig. 4a, higher $\sigma_r$ can be induced by increasing the number of cationic elements in high-entropy rare-earth hafnates (HEREHs). This results in greater lattice distortion, contributing to higher ionic disorder, elemental inhomogeneity, phase separation tendency, etc.[33,34] To explore the potential of LSS in tackling this challenge, we performed trials in incorporating a great number of elements in A and B sites of HEREHs. The XRD in Fig. S12 confirms the successful syntheses of single-phase HEREH samples with up to 20 cationic elements (15 RE-cations in the A site and 5 TM-cations in the B site). The EDS maps (see Fig. 4b) also exhibit a nanoscale homogenous distribution of each cationic element in the designed atomic ratio (Table S5) throughout the samples, indicating greatly expanded compositions of HEREHs via LSS. In addition, the performance of LSS was further compared to conventional ultrafast high-temperature synthesis methods to show its advantages. Taking the X2-monosilicate phase as an example, we sequentially incorporated 8, 10, and 15 cationic elements into its lattice. As the XRD results depicted in Fig. S13, all the samples produced by LSS adopt the pure X2-$RE_2SiO_5$ phase, whereas the samples produced by ultrafast high-temperature synthesis contain the $RE_2O_3$ and $RE_4Si_3O_{12}$ impurity. Such a result highlights a greater compositional diversity achieved through LSS. Regarding the unique G-type HEREDs, a constituent element number of up to 20 was achieved to form a pure phase, as evidenced by the XRD patterns (Fig. S14) and the EDS mapping results (Fig. 4c and Table S6). Based on these results, it is affirmed that LSS can not only realize the synthesis of certain "unsynthesizable" phases of HEREDs but also significantly expand their compositional landscape, potentially leading to unexpected properties.

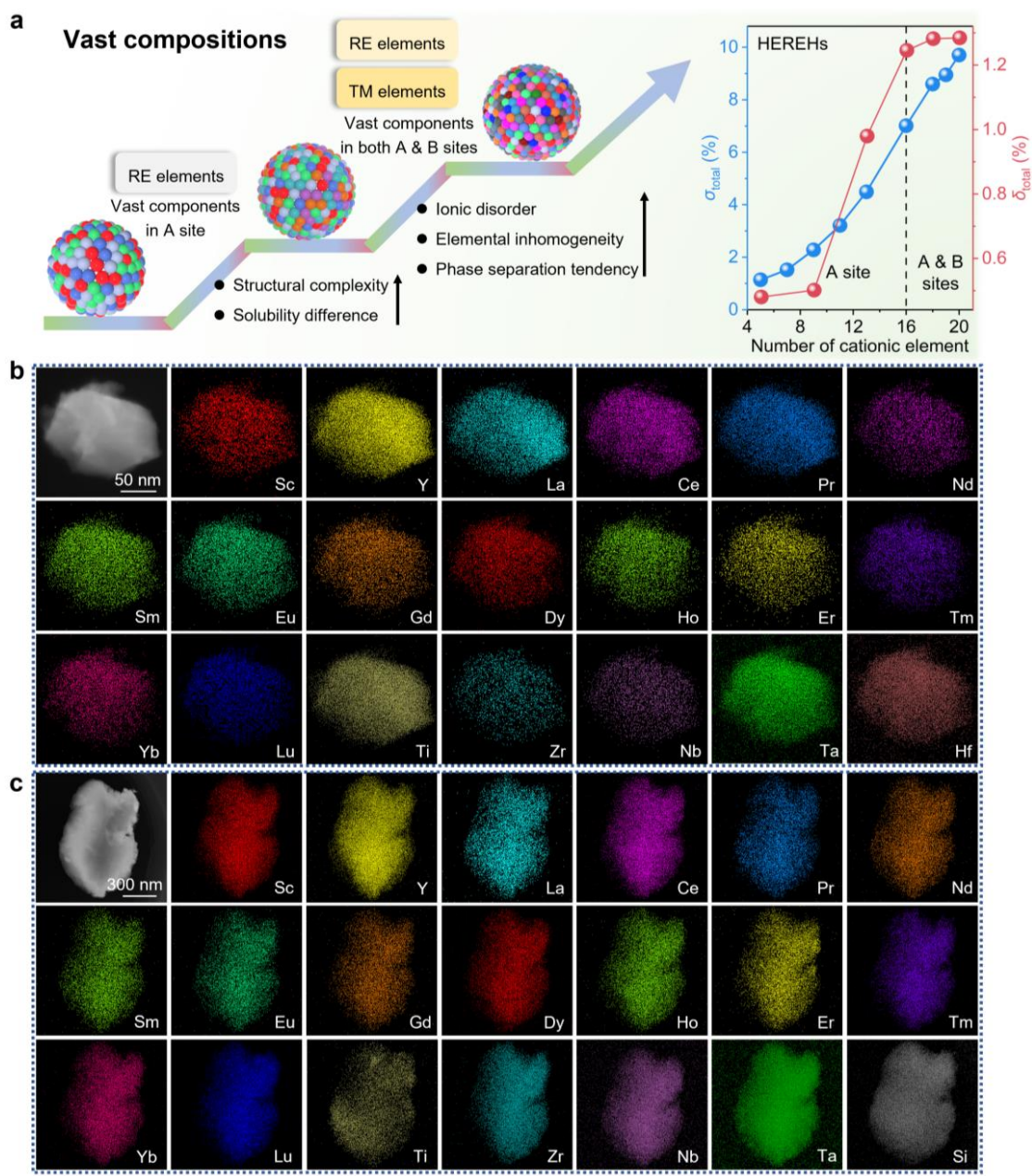

**Fig. 4 Vast compositions of HEOs achieved through LSS.** (**a**) Synthesis difficulties for HEOs with increasing constituent elements. EDS maps of the as-synthesized HEREHs (**b**) and HEREDs (**c**) with 15 RE-cationic elements in the A site and 5 TM-cationic elements in the B site.

Previous studies have demonstrated that severe lattice distortion and broadened composition range are valuable features in microwave absorption.[35] Thus, the microwave absorption performance of pure-phase HEREDs was explored. As shown in Fig. S15 and Fig. S16, a series of HERED samples were synthesized via our self-developed LSS and conventional solid-state reaction for comparison. For the $Yb_2Si_2O_7$ and 5-cation HERED samples (denoted as 1-β and 5-β, respectively) produced by solid-state reaction, almost no microwave absorption can be found in them (see Fig. S17), with a maximum RL value of only -4.6 dB and -10.0 dB (i.e., EAB of 0 GHz and 0.02 GHz, respectively). Such inferior microwave absorption performance is predominantly due to the inherently low dielectric loss characteristic of RE silicates, which makes them typical microwave-transparent agents.[36] With respect to the samples synthesized through LSS, Fig. 5a summarizes all their EAB values (raw data can be found in Fig. S18). Evidently, the microwave absorption of 1-β as well as α-, A-, β-, and γ-type 5-cation HEREDs (marked as 5-α, 5-A, 5-β, and 5-γ, respectively) remain poor, with limited EAB values of less than 0.3 GHz. In contrast, significant enhancements are achieved in δ-, F-, and G-type 5-cation HEREDs (labeled as 5-δ, 5-F, and 5-G, respectively), reaching EAB values of 0.9 GHz, 2.2 GHz, and 2.5 GHz, respectively. Notably, by incorporating 20 constituent elements in G-type HEREDs (20-G), the EAB can be further elevated to 4.3 GHz. This EAB value is 430 times higher than that of 1-β samples, representing a significant breakthrough in the microwave absorption performance of HEREDs. Furthermore, Fig. 5b displays the RL diagram of representative 20-G HERED samples, which exhibits that their EAB covers from 12.9 GHz to 17.2 GHz. Additionally, multiple semicircles can be found in the Cole-Cole diagram, as presented in Fig. 5c. These semicircles provide strong evidence for the frequent occurrence of Debye relaxation, giving rise to considerable energy dissipation of microwaves through polarization loss.

Since polarization loss is commonly associated with atomic vacancies,[37] we further carried out X-ray photoelectron spectroscopy (XPS) analysis to verify their presence. It is noteworthy that, to ensure comparability, $Yb_2Si_2O_7$ is selected due to its synthesizability with both LSS and conventional solid-solution reaction methods. The XPS results (Fig. 5d) indicate that, compared to the negligible O vacancies in the samples synthesized by solid-solution reaction, the pronounced peak at 532.3 eV demonstrates the existence of abundant O vacancies in the samples synthesized through LSS. Such a significant difference validates that LSS with its high energy density is beneficial to the formation of vacancies. In terms of O vacancy concentrations in various types of HEREDs, EPR patterns in Fig. 5e suggest that the low-temperature phases A and α possess relatively low levels of O vacancies, while the high-temperature phases β, γ, δ, F, and G show a gradual increase in vacancy concentrations. Moreover, the increase in constituent elements (from 5-G to 20-G) also greatly enhances O vacancy concentrations.

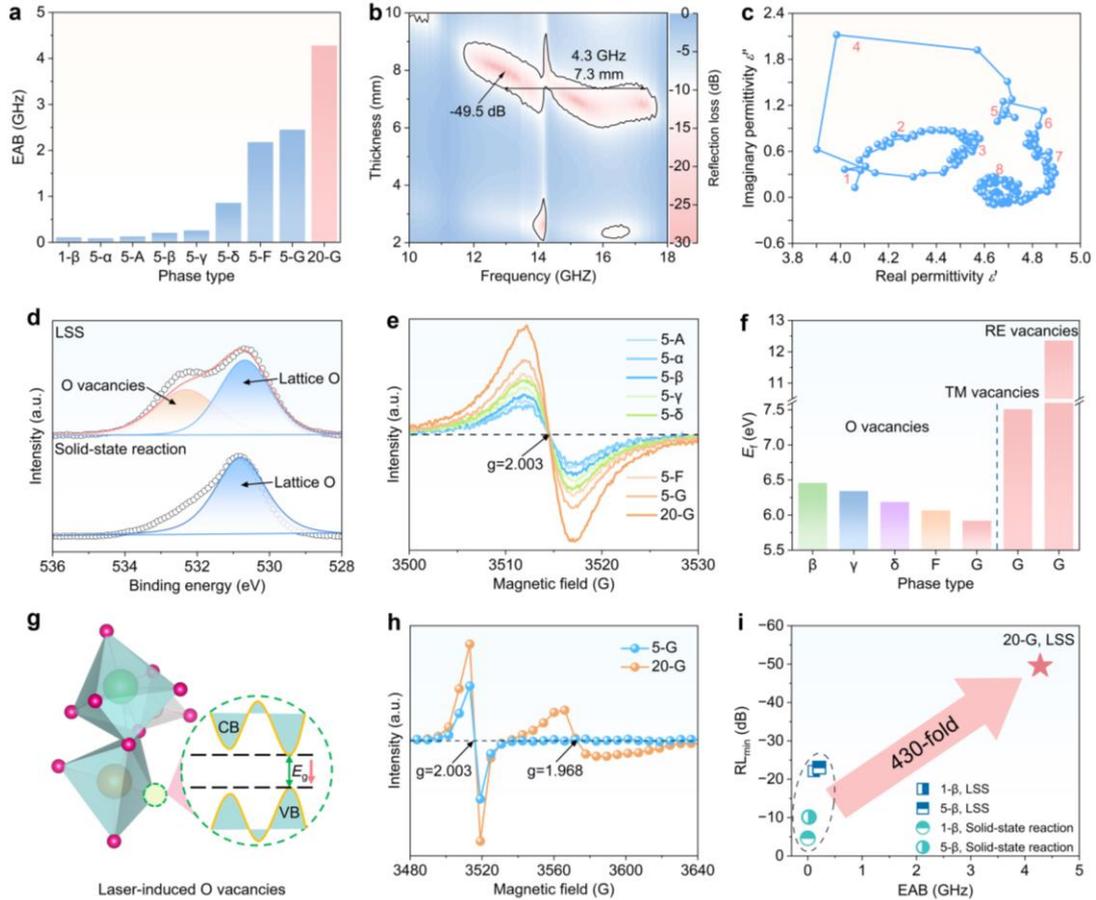

**Fig. 5 Microwave absorption of the as-synthesized HEREDs.** (**a**) EAB values for different types of HERED samples. RL diagram (**b**) and Cole-Cole diagram with real permittivity ε′ and imaginary permittivity ε″ (**c**) of representative 20-G HERED samples. (**d**) XPS patterns of representative $Yb_2Si_2O_7$ samples synthesized by LSS and solid-state reaction. (**e**) EPR spectra of all HERED samples in a range of 3500-3530 G. (**f**) Computed $E_f$ of O, TM, and RE vacancies. (**g**) Schematic of a narrowed bandgap between the conduction band (CB) and the valence band (VB) due to O vacancies. (**h**) EPR spectra of representative 5-G and 20-G HERED samples in a range of 3480-3640 G. (**i**) Comparison of microwave absorption properties for HEREDs synthesized by different methods.

To unveil the underlying mechanism of vacancies using LSS, we computed the formation energy ($E_f$) of different atomic vacancies in HEREDs. As depicted in Fig. 5f, the results demonstrate that the formation of O vacancies is the most preferable, in contrast to TM and RE vacancies. For $E_f$ of O vacancies in all HEREDs, β-phase has the highest value, implying that O vacancies are the hardest to form in β-type HEREDs. Conversely, the $E_f$ value for the G-phase HERED is the lowest, aligning perfectly with the EPR results presented in Fig. 5e. Since the rich O vacancies generated in G-HEREDs can amplify polarization loss by inducing numerous dipoles in their vicinity, it helps to facilitate energy dissipation of interacting microwaves through the frequent reorientation of these dipoles. Besides polarization loss, as previously evidenced in Fig. S5, the formation of O vacancies can also narrow the bandgap of G-HEREDs (see Fig. 5g), contributing to enhanced conduction loss. In addition, the additional signal in Fig. 5h found at g=1.968 for the as-synthesized 20-G HERED samples may be ascribed to certain TM vacancies.[38] All these phenomena verify severe lattice distortion in 20-G HEREDs due to the incorporation of a large number of cationic elements, resulting in exacerbated polarization and conduction losses, and, thereby, a significant breakthrough of microwave absorption with remarkable 430-fold improvement in EAB (see Fig. 5i).

**Conclusions**

In summary, we have developed a universal and effective LSS technique that enables high-throughput synthesis of a series of HEOs with various crystal structures, including rock-salt, perovskite, spinel, fluorite, pyrochlore, tantalate, and silicate, incorporating up to 20 cationic elements. In particular, the high-energy-density laser not only achieves an ultrahigh synthesis temperature of up to ~3500 °C within milliseconds but also promotes the formation of unique cavitation bubbles to inhibit the evaporation of low-melting raw oxides. These distinctive synthesis characteristics facilitate the complete mixing of immiscible elements in complex crystal structures, allowing for the creation of all seven types of HERED phases, including unprecedented δ-, F-, and G-type structures that have never been reported in HEREDs previously. Beneficial from the unique G-type structure and the 20-cation composition that causes severe lattice distortion, the as-synthesized 20-G samples with abundant TM and O atomic vacancies are granted a new functionality of microwave absorption with an EAB of up to 4.3 GHz. The as-developed general LSS approach opens a vast structural and compositional space for HEOs, as well as potentially bestowing them with new properties that are highly valuable for various applications.

## Methods

### Sample synthesis

The commercially available oxide powders, including RE oxides of $Sc_2O_3$, $Y_2O_3$, $La_2O_3$, $CeO_2$, $Pr_6O_{11}$, $Nd_2O_3$, $Sm_2O_3$, $Eu_2O_3$, $Gd_2O_3$, $Dy_2O_3$, $Ho_2O_3$, $Er_2O_3$, $Tm_2O_3$, $Yb_2O_3$, and $Lu_2O_3$ (Purity: 99.9%, particle size: 1-3 μm, Shanghai Maclin Biochemical Technology Co. Ltd., Shanghai, China), TM oxides of $TiO_2$, $Cr_2O_3$, $MnO_2$, $Fe_2O_3$, CoO, NiO, CuO, $ZrO_2$, $Nb_2O_5$, $HfO_2$, and $Ta_2O_5$ (Purity: 99.9%, particle size: 1-3 μm, Shanghai Maclin Biochemical Technology Co. Ltd., Shanghai, China), $Al_2O_3$ (Purity: analytical pure, particle size: >100 mesh, Sinopharm Chemical Reagent Co., Ltd., Shanghai, China), and $SiO_2$ (Purity: 99.9%, particle size: 2 μm, Shanghai Maclin Biochemical Technology Co. Ltd., Shanghai, China) were utilized as raw materials. First, the commercially available oxide powders were weighed and mixed in an agate mortar according to the stoichiometric ratio of the reaction equation. The as-obtained mixtures were then ball-milled using a high-throughput ball-milling apparatus for 6 h and dried at 100 °C for 4 h, followed by being moved into the graphite crucible with a diameter of 10 mm and uniformly compacted.

Subsequently, the LSS apparatus was applied to synthesize HEO powders. Specifically, a fiber continuous 1080 nm-wave laser (RFL-C3000S, Raycus Fiber Laser Technology Co., Ltd., Beijing, China) was applied as the energy source. The laser beam diameter was adjusted to 10 mm through a focusing lens to ensure uniform energy distribution across the entire synthetic area. The system parameters, including powers (1000-1500 W) and pulse duty cycle modulation (20%-100%), were tuned according to the experimental requirements. An infrared thermometer (M313, SensorTherm GmbH, Bavaria, Germany) and an infrared thermal camera (MAG-HT60ET, Magnity

Technologies, Shanghai, China) were used for temperature measurement and thermal imaging. A high-speed camera (sh6-116-m-4t, Chengdu Libo Optoelectronics Technology Co., Ltd., Changchun, China) with a frame rate of 15800 FPS, equipped with a band-pass filter, was employed for real-time observations. Regarding the samples for microwave absorption testing, the as-produced HERED powders were uniformly blended with paraffin in a weight ratio of 3:7 at 90 °C for 1 h. After natural cooling, the samples were machined into concentric rings with an inner diameter of 3.04 mm, an outer diameter of 7.00 mm, and a thickness of 3 mm for the measurement of electromagnetic parameters.

**Sample characterization**

The phase composition of as-produced HEO samples was detected by XRD (X'pert PRO, PANalytical, Almelo, Netherlands), and the corresponding Rietveld refinement was performed with the GSAS software.[39] The morphology, atomic structure, and compositional uniformity were characterized by an aberration-corrected STEM (HF5000, Hitachi High-Tech Co., Ltd., Tokyo, Japan) equipped with EDS. The absorption spectra were measured by a UV-Vis-NIR spectrophotometer (Lamda-950, Perkin Elmer, Massachusetts, USA). The absorbance $A$ of samples was derived from the modified Lambert-Beer law as expressed by:[40]

$$A = -log_{10} R \tag{1}$$

where $R$ represents the reflection parameter. The optical bandgap ($E_g$) was determined through Tauc analysis of UV-Vis-NIR diffuse reflectance spectra:[41]

$$(F(R)h\nu)^{1/2} = \frac{(1-R)^2}{2R} = B(h\nu - E_g) \tag{2}$$

where $F(R)$ is the Kubelka-Munk function, $h\nu$ is photon energy, and $B$ is the Tauc constant. The $E_g$ values were obtained from the plot of the Tauc equation, whereby the intersection of the tangent to the turning point with the x-axis was used for the determination. In addition, the electromagnetic parameters of the samples were measured by a vector network analyzer (VNA, E5071C, Agilent Technologies, California, USA) in the range of 2-18 GHz. The RL values of the samples were calculated based on the transmission line theory as follows:[35]

$$RL = 20log \left|\frac{Z_{in}-Z_0}{Z_{in}+Z_0}\right| \quad (3)$$

$$Z_{in} = Z_0\sqrt{\frac{\mu_r}{\varepsilon_r}}tanh\left[j(\frac{2\pi fd}{c})\sqrt{\mu_r\varepsilon_r}\right] \quad (4)$$

where $Z_{in}$ and $Z_0$ are the normalized input impedance and the impedance of free space, respectively. $\varepsilon_r$ and $\mu_r$ are the complex permittivity and permeability, respectively. $f$, $c$, and $d$ are the incident microwave frequency, speed of light, and thickness of the tested sample, respectively. The vacancies of samples were identified by XPS (Escalab 250Xi, Thermo Fisher Scientific, Massachusetts, USA) and EPR (A300, Bruker, Karlsruhe, Germany).

**Calculation**

The average ion radius $\bar{r}$ and the difference of ion radii $\sigma_r$ were used as the descriptor for determining the synthesizability of HEREDs, which were calculated by the following formulae:[13]

$$\bar{r} = \sum_{i=1}^{n} c_i r_i \quad (5)$$

$$\sigma_r = \sqrt{\sum_{i=1}^{n} c_i(1-\frac{r_i}{\bar{r}})^2} \times 100\% \quad (6)$$

where $n$ is the number of RE element species, $c_i$ and $r_i$ are the molar fraction and the ion radii of the $i$-th RE element in the sublattice. Specifically, for the presence of two cation sites, the total difference $\sigma_{total}$ of ion radii was computed as follows:[5]

$$\sigma_{total} = \sqrt{\sigma_{r_A}^2 + \sigma_{r_B}^2} \quad (7)$$

First-principles calculations were conducted via the Vienna Ab initio Simulation Package.[42] The projector-augmented wave (PAW) method was employed to describe the interaction between the core and valence electrons.[43] The exchange-correlation interactions were treated within the generalized gradient approximation (GGA)[43] using the Perdew-Burke-Ernzerhof functional (PBE).[45] A plane-wave basis set with an energy cutoff of 400 eV was used to expand the wave functions. Supercells (88, 176, 176, 264, 290, 290, and 290 atoms for 4-cation HEREDs and 4-, 8-, 12-, 15-, 18-, and 20-cation HEREHs, respectively) were constructed based on the primitive unit cells of each phase of HEREDs and HEREHs, and the special quasi-random structure approach, implemented in the Alloy Theoretic Automated Toolkit[46] was applied to generate the disordered occupations of the metal atoms for high-entropy models. Oxygen and metal vacancies were introduced by removing one oxygen or metal atom from the respective sites in the models. The Brillouin zone was sampled employing the $\Gamma$-centered method with a separation of approximately 0.3 Å$^{-1}$.[47] Both the defect-free and defect-containing supercells were fully relaxed to their equilibrium geometries. The convergence criteria for the electronic self-consistent loop were set to 10$^{-5}$ eV. The $E_f$ was calculated using the formula:[48]

$$E_f = E_{def} - E_{perfect} + \sum_i^n n_i \mu_i \quad (8)$$

where $E_{def}$ is the total energy of the supercell containing the defect, $E_{perfect}$ is the total energy of the perfect supercell, $n_i$ is the number of vacancy atoms of species $i$, and $\mu_i$ is the chemical potential of species. For oxygen vacancies, the chemical potential of $\mu_O$ was taken as half the total energy of an O$_2$ molecule in the gas phase. For metal vacancies, the chemical potentials of the respective metals were referenced to the energy of the metal in its standard state. The lattice distortion degree of $\delta_{[AO_n]}$ and $\delta_{[BO_n]}$ polyhedron in HERHs can be obtained as follows, respectively:[49]

$$\delta_{[AO_n]} = \frac{1}{n}\sum_{i=1}^{n}(1 - \frac{d_{Ai}}{\overline{d_A}})^2 \times 100\% \quad (9)$$

$$\delta_{[BO_n]} = \frac{1}{n}\sum_{i=1}^{n}(1 - \frac{d_{Bi}}{\overline{d_B}})^2 \times 100\% \quad (10)$$

Here, $d_{A_i}$ represents the distances of the A-O bond between the $i$-th O atom and the A atom, and $\overline{d_A}$ represents the average A-O distance in the [AO$_n$] polyhedron. Similarly, $d_{B_i}$ represents the distances of the B-O bond between the $i$-th O atom and the A atom, and $\overline{d_B}$ represents the average B-O distance in the [BO$_n$] polyhedron. Therefore, the total degree of lattice distortion can be expressed as follows:

$$\delta_{total} = \sqrt{\delta_{[AO_n]}^2 + \delta_{[BO_n]}^2} \quad (11)$$

# References

1. C. Rost, E. Sachet, T. Borman, A. Moballegh, E. Dickey, D. Hou, J. Jones, S. Curtarolo, J. Maria, Entropy-stabilized oxides, *Nat. Commun.* **6**, 8485 (2015).

2. A. Sarkar, Q. Wang, A. Schiele, M. Chellali, S. Bhattacharya, D. Wang, T. Brezesinski, H. Hahn, L. Velasco, B. Breitung, High-entropy oxides: fundamental aspects and electrochemical properties, *Adv. Mater.* **31**, 1806236 (2019).

3. C. Oses, C. Toher, S. Curtarolo, High-entropy ceramics, *Nat. Rev. Mater.* **5**, 295-309 (2020).

4. A. Sarkar, L. Velasco, D. Wang, Q. Wang, G. Talasila, L. Biasi, C. Kübel, T. Brezesinski, S. Bhattacharya, H. Hahn, B. Breitung, High entropy oxides for reversible energy storage, *Nat. Commun.* **9**, 3400 (2018).

5. S. Jiang, T. Hu, J. Gild, N. Zhou, J. Nie, M. Qin, T. Harrington, K. Vecchio, J. Luo, A new class of high-entropy perovskite oxides, *Scripta Mater.* **142**, 116-120 (2018).

6. M. Folgueras, Y. Jiang, J. Jin, P. Yang, High-entropy halide perovskite single crystals stabilized by mild chemistry, *Nature* **621**, 282-288 (2023).

7. L. Su, H. Huyan, A. Sarkar, W. Gao, X. Yan, C. Addiego, R. Kruk, H. Hahn, X. Pan, Direct observation of elemental fluctuation and oxygen octahedral distortion-dependent charge distribution in high entropy oxides, *Nat. Commun.* **13**, 2358 (2022).

8. J. Dąbrowa, M. Stygar, A. Mikuła, A. Knapik, K. Mroczka, W. Tejchman, M. Danielewski, M. Martin, Synthesis and microstructure of the (Co, Cr, Fe, Mn, Ni)$_3$O$_4$ high entropy oxide characterized by spinel structure, *Mater. Lett.* **216**, 32-36 (2018).

9. J. Gild, M. Samiee, J. Braun, T. Harrington, H. Vega, P. Hopkins, K. Vecchio, J. Luo, High-entropy fluorite oxides, *J. Eur. Ceram. Soc.* **38**, 3578-3584 (2018).

10. K. Chen, X. Pei, L. Tang, H. Cheng, Z. Li, C. Li, X. Zhang, L. An, A five-component entropy-stabilized fluorite oxide, *J. Eur. Ceram. Soc.* **38**, 4161-4164 (2018).

11. A. Wright, Q. Wang, S. Ko, K. Chung, R. Chen, J. Luo, Size disorder as a descriptor for predicting reduced thermal conductivity in medium- and high-entropy pyrochlore oxides, *Scripta Mater.* **181**, 76-81 (2020).

12. Y. Dong, K. Ren, Y. Lu, Q. Wang, J. Liu, Y. Wang, High-entropy environmental barrier coating for the ceramic matrix composites, *J. Eur. Ceram. Soc.* **39**, 2574-2579 (2019).

13. Y. Luo, L. Sun, J. Wang, T. Du, C. Zhou, J. Zhang, J. Wang, Phase formation capability and compositional design of β-phase multiple rare-earth principal component disilicates, *Nat. Commun.* **14**, 1275 (2023).

14. Y. Li, X. Bai, D. Yuan, C. Yu, X. San, Y. Guo, L. Zhang, J. Ye, Cu-based high-entropy two-dimensional oxide as stable and active photothermal catalyst, *Nat. Commun.* **14**, 3171 (2023).

15. Y. Zeng, B. Ouyang, J. Liu, Y. Byeon, Z. Cai, L. Miara, Y. Wang, G. Ceder, High-entropy mechanism to boost ionic conductivity, *Science* **378**, 1320-1324 (2022).

16. X. Wang, M. Cheng, G. Xiao, C. Wang, R. Qiao, F. Zhang, Y. Bai, Y. Li, Y. Wu, Z. Wang, Preparation and corrosion resistance of high-entropy disilicate $(Y_{0.25}Yb_{0.25}Er_{0.25}Sc_{0.25})_2Si_2O_7$ ceramics, *Corros. Sci.* **192**, 109786 (2021).

17. F. Ding, C. Zhao, D. Xiao, X. Rong, H. Wang, Y. Li, Y. Yang, Y. Lu, Y. Hu, Using high-entropy configuration strategy to design Na-ion layered oxide cathodes with superior electrochemical performance and thermal stability, *J. Am. Chem. Soc.* **144**, 8286-8295 (2022).

18. B. Yang, Y. Zhang, H. Pan, W. Si, Q. Zhang, Z. Shen, Y. Yu, S. Lan, F. Meng, Y. Liu, H. Huang, J. He, L. Gu, S. Zhang, L. Chen, J. Zhu, C. Nan, Y. Lin, High-entropy enhanced capacitive energy storage, *Nat. Mater.* **21**, 1074-1080 (2022).

19. B. Jiang, W. Wang, S. Liu, Y. Wang, C. Wang, Y. Chen, L. Xie, M. Huang, J. He, High figure-of-merit and power generation in high-entropy GeTe-based thermoelectrics, *Science* **377**, 208-213 (2022).

20. A. Sarkar, R. Djenadic, N. Usharani, K. Sanghvi, V. Chakravadhanula, A. Gandhi, H. Hahn, S. Bhattacharya, Nanocrystalline multicomponent entropy stabilised transition metal oxides, *J. Eur. Ceram. Soc.* **37**, 747-754 (2017).

21. D. Wang, Z. Liu, S. Du, Y. Zhang, H. Li, Z. Xiao, Wei Chen, R. Chen, Y. Wang, Y. Zou, S. Wang, Low-temperature synthesis of small-sized high-entropy oxides for water oxidation, *J. Mater. Chem. A* **7**, 24211-24216 (2019).

22. S. Liu, C. Pao, J. Chen, S. Li, K. Chen, Z. Xuan, C. Song, J. Urban, M. Swihart, C. Dun, A general flame aerosol route to high-entropy nanoceramics, *Matter* **7**, 3994-4013 (2024).

23. T. Li, Y. Yao, Z. Huang, P. Xie, Z. Liu, M. Yang, J. Gao, K. Zeng, A. Brozena, G. Pastel, M. Jiao, Q. Dong, J. Dai, S. Li, H. Zong, M. Chi, J. Luo, Y. Mo, G. Wang, C. Wang, R. Yassar, L. Hu, Denary oxide nanoparticles as highly stable catalysts for methane combustion, *Nat. Catal.* **4**, 62-70 (2021).

24. S. Barcikowski, A. Plech, K. Suslick, A. Vogel, Materials synthesis in a bubble, *MRS Bull.* **44**, 382-391 (2019).

25. Y. Li, X. Liu, T. Wu, X. Zhang, H. Han, X. Liu, Y. Chen, Z. Tang, Z. Liu, Y. Zhang, H. Liu, L. Zhao, D. Ma, W. Zhou, Pulsed laser induced plasma and thermal effects on molybdenum carbide for dry reforming of methane, *Nat. Commun.* **15**, 5495 (2024).

26. B. Wang, C. Wang, X. Yu, Y. Cao, L. Gao, C. Wu, Y. Yao, Z. Lin, Z. Zou, General synthesis of high-entropy alloy and ceramic nanoparticles in nanoseconds, *Nat. Synth.* **1**, 138-146 (2022).

27. I. Bondar, Rare-earth silicates, *Ceram. Int.* **8**, 83-89 (1982).

28. P. Chen, Y. Zhou, J. He, W. Jiang, J. Li, H. Ni, Q. Zhang, L. Lin, On the crystal chemistry of $RE_2Si_2O_7$: Revisited structures, group-subgroup relationship, and insights of $Ce^{3+}$-activated radioluminescence, *Chem. Mater.* **35**, 2635-2646 (2023).

29. Y. Liu, J. Yuan, J. Zhou, K. Pan, R. Zhang, R. Zhao, L. Li, Y. Huang, Z. Liu. Laser solid-phase synthesis of graphene shell-encapsulated high-entropy alloy nanoparticles. *Light Sci. Appl.* **13**, 270 (2024).

30. X. Chen, C. Meng, Y. Wang, Q. Zhao, Y. Li, X. Chen, D. Yang, Y. Li, Y. Zhou. Laser-synthesized rutile $TiO_2$ with abundant oxygen vacancies for enhanced solar water evaporation, *ACS Sustainable Chem. Eng.* **8**, 1095-1101 (2020).

31. Z. Li, M. Huang, B. Chan, J. Ge, D. Xin, D. Jiang, H. Liu, W. Zhou, Laser constructed bulk oxygen vacancy caused high P doping for boosting the sodium-storage capability, *Interdiscip. Mater.* **2**, 876-887 (2023).

32. L. Lopato, Highly refractory oxide systems containing oxides of rare-earth elements, *Ceram. Int.* **2**, 18-32 (1976).

33. Y. Liao, Y. Li, R. Zhao, J. Zhang, L. Zhao, L. Ji, Z. Zhang, X. Liu, G. Qin, X. Zhang, High-entropy-alloy nanoparticles with 21 ultra-mixed elements for efficient photothermal conversion, *Natl. Sci. Rev.* **9**, 41 (2022).

34. G. Cao, J. Liang, Z. Guo, K. Yang, G. Wang, H. Wang, X. Wan, Z. Li, Y. Bai, Y. Zhang, J. Liu, Y. Feng, Z. Zheng, C. Lu, G. He, Z. Xiong, Z. Liu, S. Chen, Y. Guo, M. Zeng, J. Lin, L. Fu, Liquid metal for high-entropy alloy nanoparticles synthesis, *Nature* **619**, 73-77 (2023).

35. F. Gu, W. Wang, H. Meng, Y. Liu, L. Zhuang, H. Yu, Y. Chu, Lattice distortion boosted exceptional electromagnetic wave absorption in high-entropy diborides, *Matter* **8**, 102004 (2025).

36. Y. Du, Z. Tian, L. Zheng, Z. Chen, K. Ming, Bin Li, Mechanical and dielectric properties of $RE_2SiO_5$ (RE=Ho, Er, Tm, Yb, and Lu) as high-temperature wave-transparent materials, *Ceram. Int.* **50**, 32187-32197 (2024).

37. B. Zhao, Z. Yan, Y. Du, L. Rao, G. Chen, Y. Wu, L. Yang, J. Zhang, L. Wu, D. Zhang, R. Che, High-entropy enhanced microwave attenuation in titanate perovskites, *Adv. Mater.* **35**, 2210243 (2023).

38. V. Grachev, G. Malovichko, Structures of impurity defects in lithium niobate and tantalate derived from electron paramagnetic and electron nuclear double resonance data, *Crystals* **11**, 339 (2021).

39. B. Toby, D. Von, GSAS-II: the genesis of a modern open-source all purpose crystallography software package, *J. Appl. Crystallogr.* **46**, 544-549 (2013).

40. W. Mäntele, Erhan Deniz, UV-Vis absorption spectroscopy: Lambert-Beer reloaded, *Spectrochim. Acta A* **173**, 965-968 (2017).

41. R. Gerdes, H. Bettentrup, D. Enseling, Markus Haase, T. Jüstel, On the synthesis, phase optimisation and luminescence of some rare earth pyrosilicates, *J. Lumin.* **190**, 451-456 (2017).

42. G. Kresse, J. Furthmüller, Efficiency of ab-initio total energy calculations for metals and semiconductors using a plane-wave basis set, *Comput. Mater. Sci.* **6**, 15-50 (1996).

43. P. Blochl, Projector augmented-wave method, *Phys. Rev. B Condens. Matter.* **50**, 17953-17979 (1994).

44. J. Perdew, K. Burke, M. Ernzerhof, Generalized gradient approximation made simple, *Phys. Rev. Lett.* **77**, 3865-3868 (1996).

45. J. Hubbard, Electron correlations in narrow energy bands, *Proc. R. Soc. Lond. A* **276**, 238-257 (1963).

46. A. Walle, P. Tiwary, M. Jong, D. Olmsted, M. Asta, A. Dick, D. Shin, Y. Wang, L. Chen, Z. Liu, Efficient stochastic generation of special quasirandom structures, *Calphad* **42**, 13-18 (2013).

47. H. Monkhorst, J. Pack, Special points for Brillouin-zone integrations. *Phys. Rev B* **13**, 5188 (1976).

48. C. Freysoldt, B. Grabowski, T. Hickel, J. Neugebauer, G. Kresse, A. Janotti, C. Van de Walle, First-principles calculations for point defects in solids, *Rev. Mod. Phys.* **86**, 253-305 (2014).

49. Y. Li, Q. Wu, M. Lai, J. Zhao, Y. Liu, Y. Fan, Y. Yao, B. Liu, Influence of chemical disorder on mechanical and thermal properties of multi-component rare earth zirconate pyrochlores $(nRE_{1/n})_2Zr_2O_7$, *J. Appl. Phys.* **132**, 075108 (2022).

**Acknowledgments**

Authors acknowledge the financial support from the National Key Research and Development Program of China (No. 2022YFB3708600), National Natural Science Foundation of China (No. 52472072), and Guangdong Basic and Applied Basic Research Foundation (2023B1515040011).

**Author contributions**

Y.C conceived and designed this work. P.W and H.B performed experiments. Y.L and H.Y performed calculations. Y.C, P.W, and L.Z analyzed the data and wrote the manuscript. All authors commented on the manuscript.

**Competing interests**

Authors declare that they have no competing interests.

# Supplementary Materials

## Laser-driven solid-state synthesis of high-entropy oxides

Peng wei, Yiwen Liu, Hao Bai, Lei Zhuang[*], Hulei Yu, Yanhui Chu[*]

School of Materials Science and Engineering, South China University of Technology; Guangzhou, 510641, China.

[*]Corresponding author. lzhuang@scut.edu.cn (L.Z); chuyh@scut.edu.cn (Y.C).

**Supplementary Materials include:**
Figs. S1 to S18
Tables S1 to S6

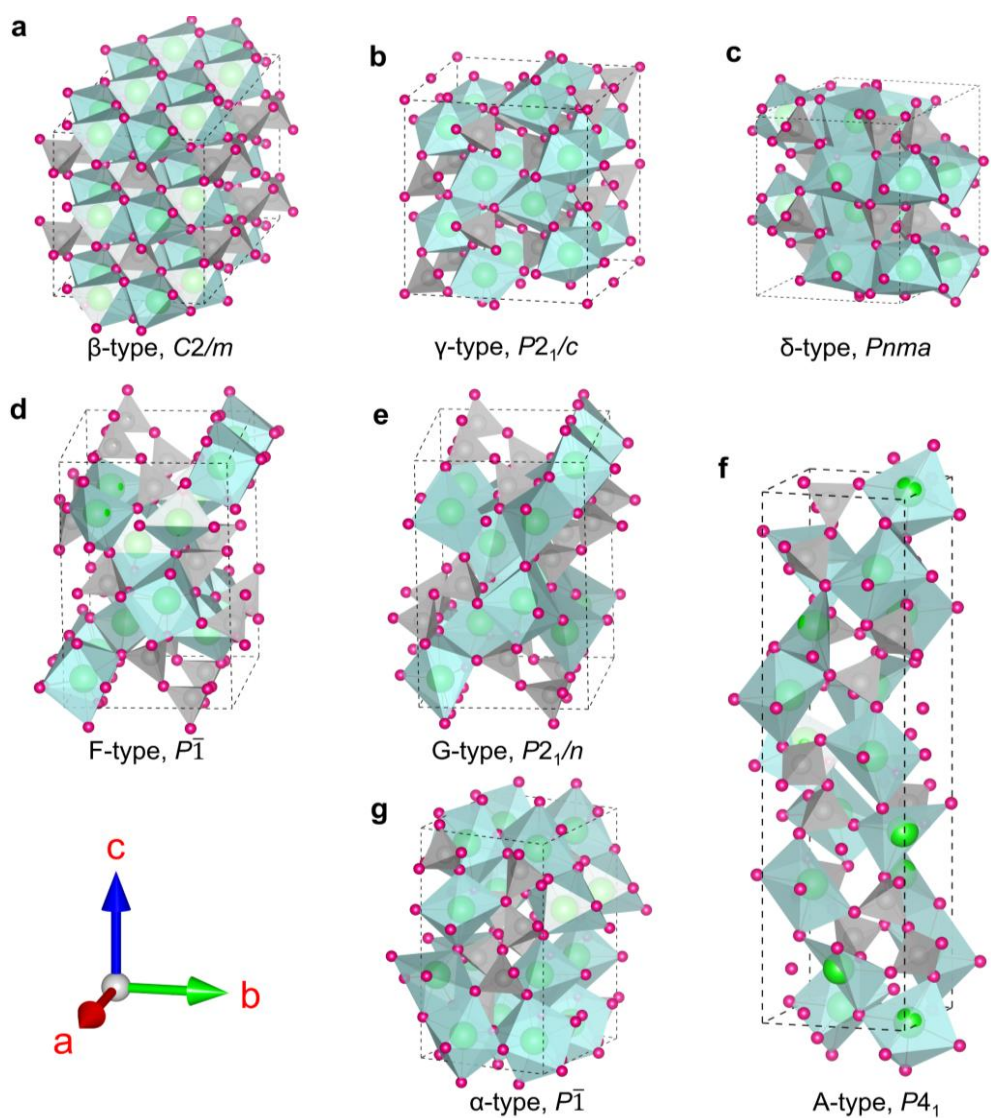

**Fig. S1** Crystal structures of all types of HEREDs. (**a**) β-type structure. (**b**) γ-type structure. (**c**) δ-type structure. (**d**) F-type structure. (**e**) G-type structure. (**f**) A-type structure. (**g**) α-type structure.

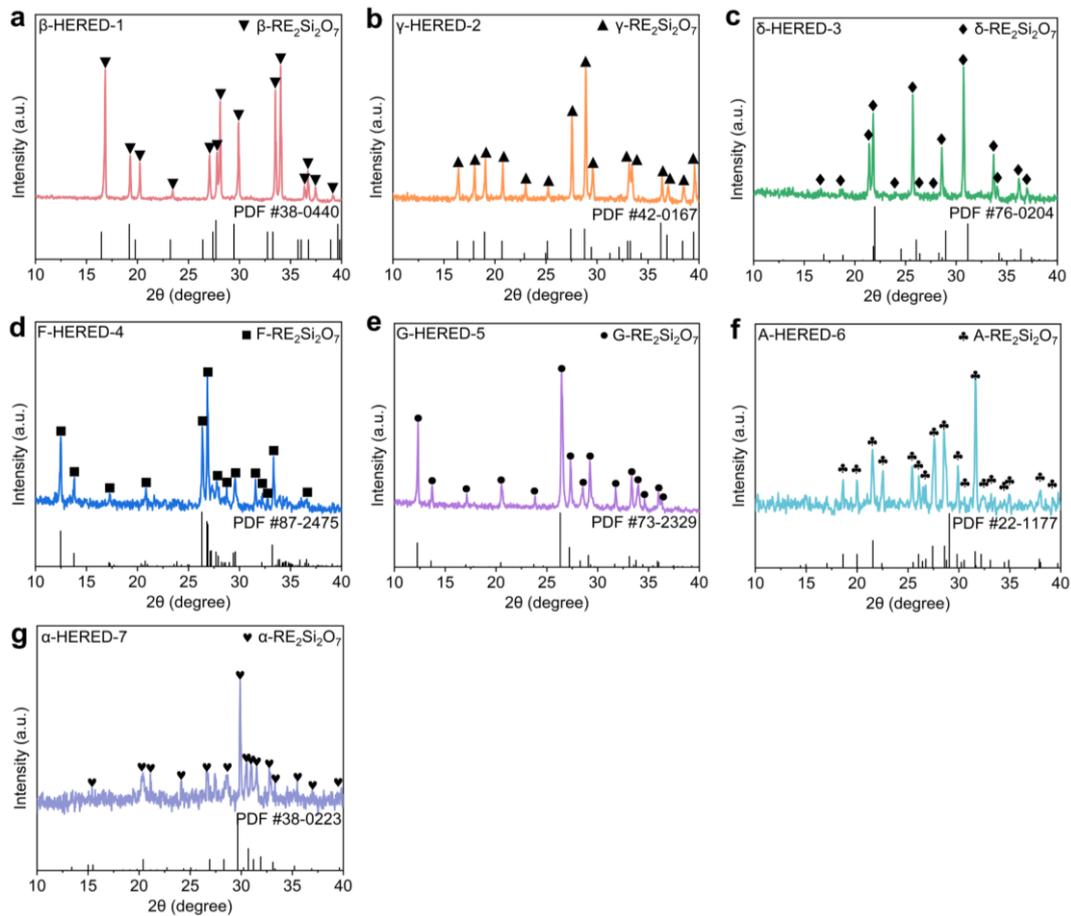

**Fig. S2** XRD patterns of the as-synthesized all types HEREDs samples. (**a**) β-HERED-1. (**b**) γ-HERED-2. (**c**) δ-HERED-3. (**d**) F-HERED-4. (**e**) G-HERED-5. (**f**) A-HERED-6. (**g**) α-HERED-7.

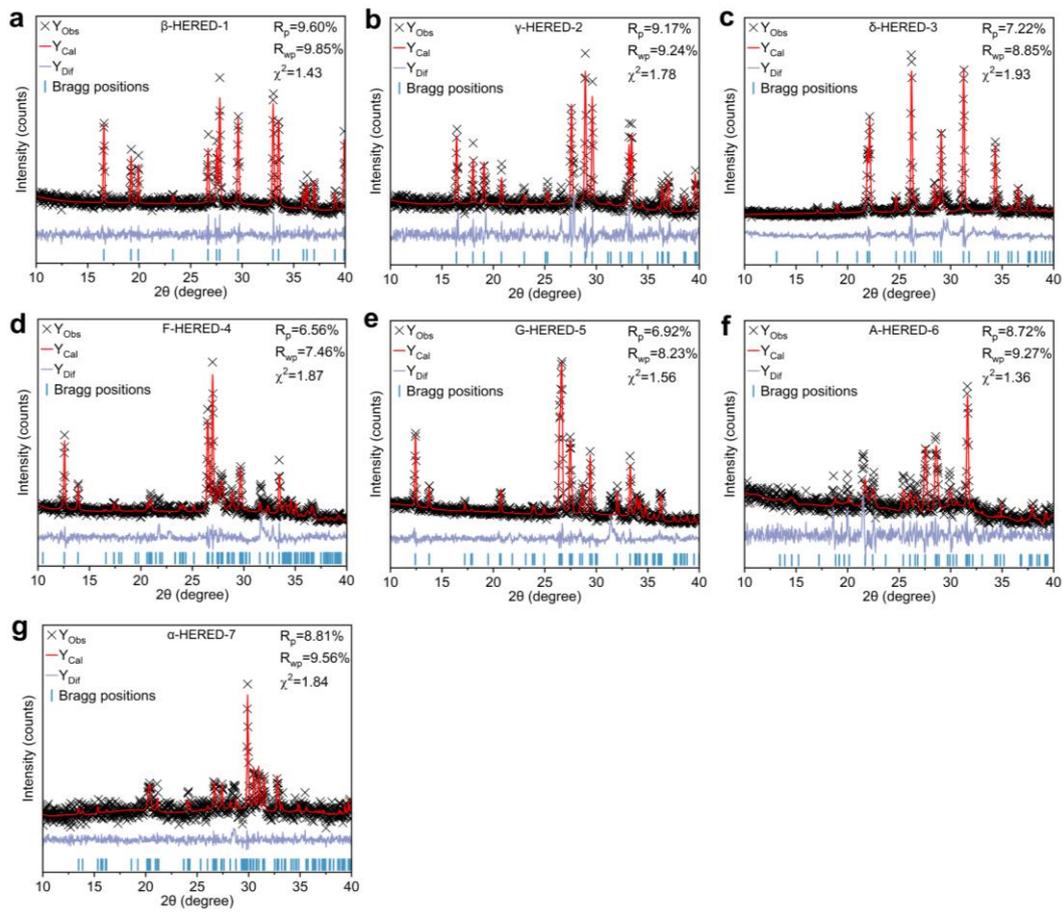

**Fig. S3** XRD Rietveld refinements of the as-synthesized all types of HEREDs samples. (**a**) β-HERED-1. (**b**) γ-HERED-2. (**c**) δ-HERED-3. (**d**) F-HERED-4. (**e**) G-HERED-5. (**f**) A-HERED-6. (**g**) α-HERED-7.

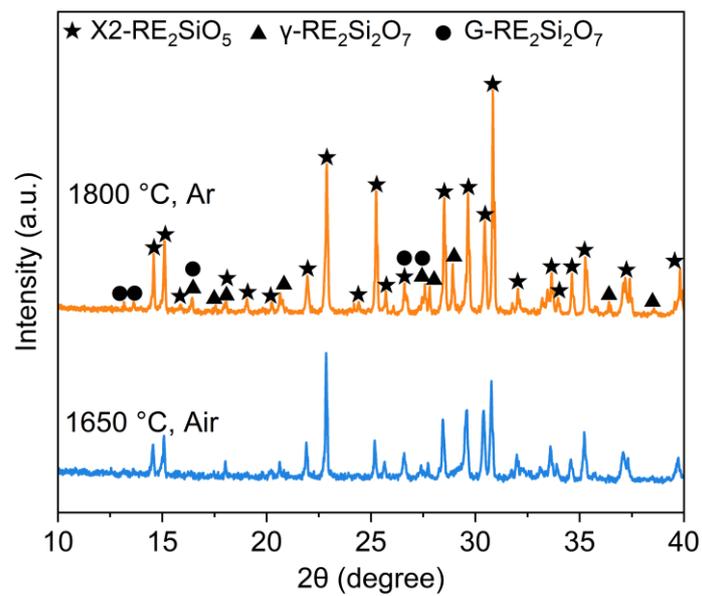

**Fig. S4** XRD patterns of the representative G-HERED-5 samples synthesized by conventional solid-state reaction at different conditions.

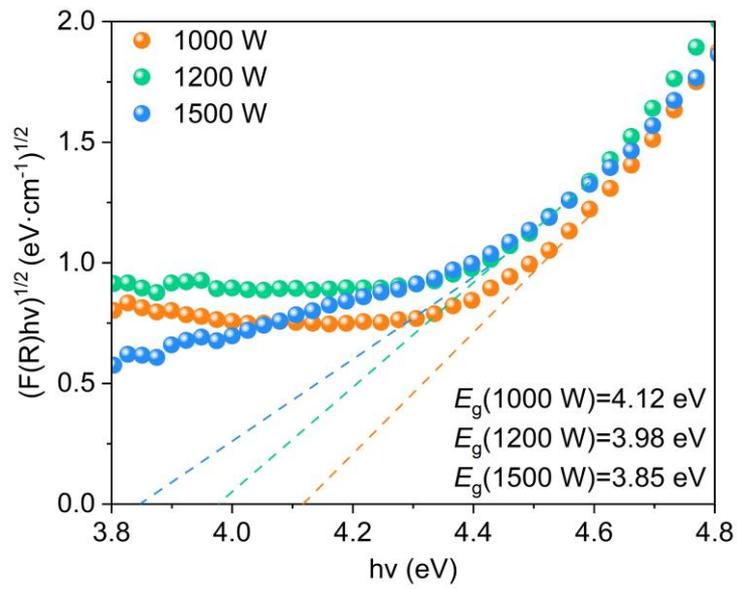

**Fig. S5** Calculated bandgaps of the representative G-HERED-5 samples synthesized at different laser power conditions from their absorbance spectra.

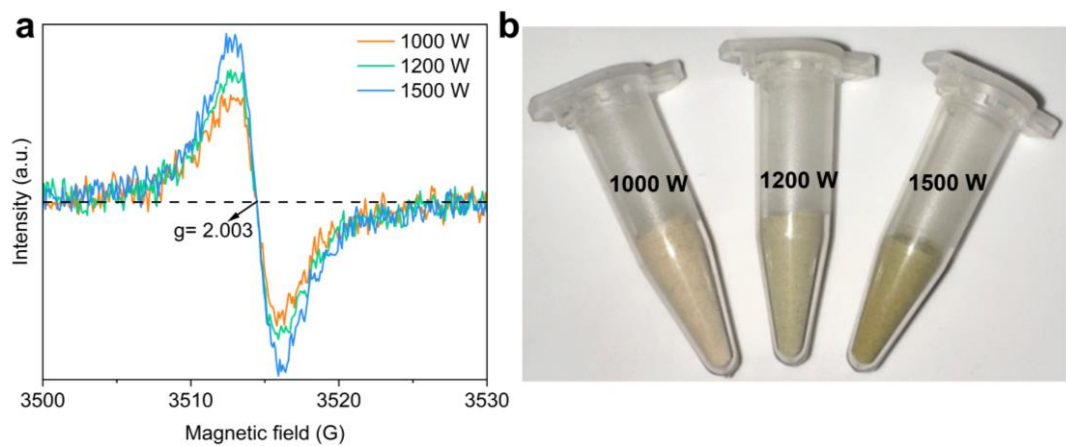

**Fig. S6** EPR patterns (**a**) and macroscopic image (**b**) of the representative G-HERED-5 samples synthesized at different laser power conditions.

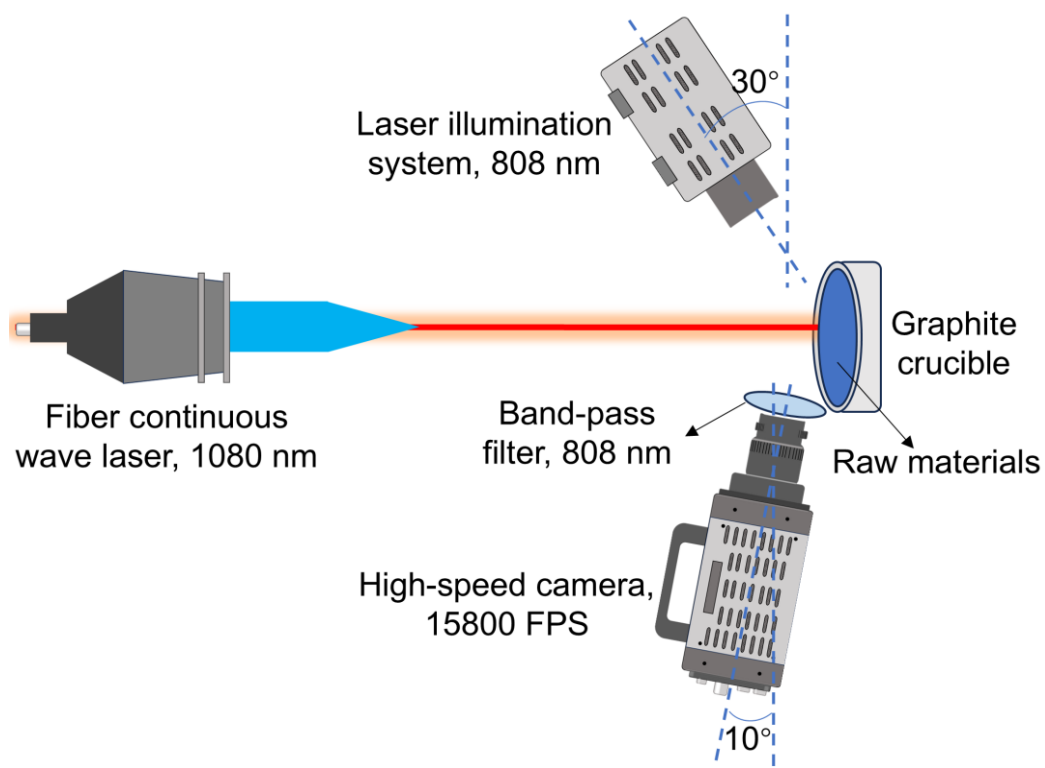

**Fig. S7** Schematic of capturing real-time images of cavitation bubbles during the synthesis process using a high-speed camera.

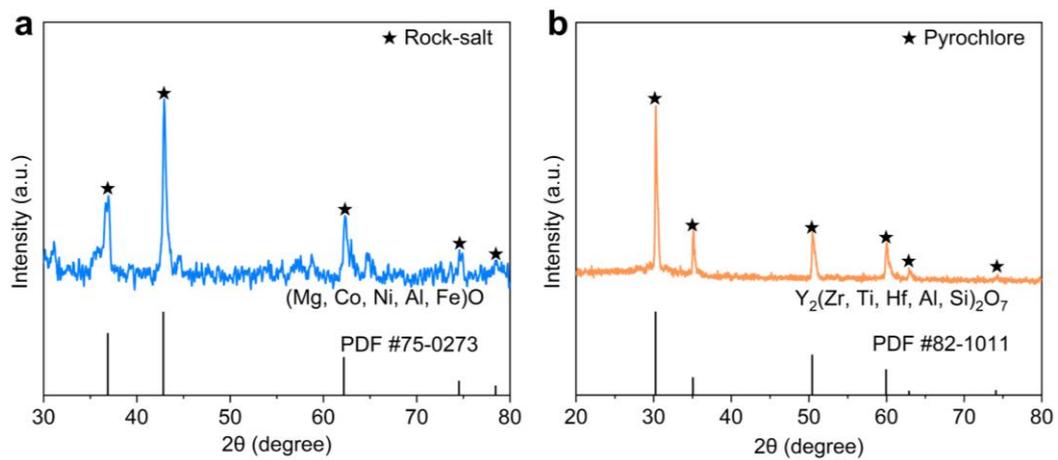

**Fig. S8** XRD patterns of the as-synthesized high-entropy pyrochlore oxide (**a**) and high-entropy rock-salt oxide (**b**) samples.

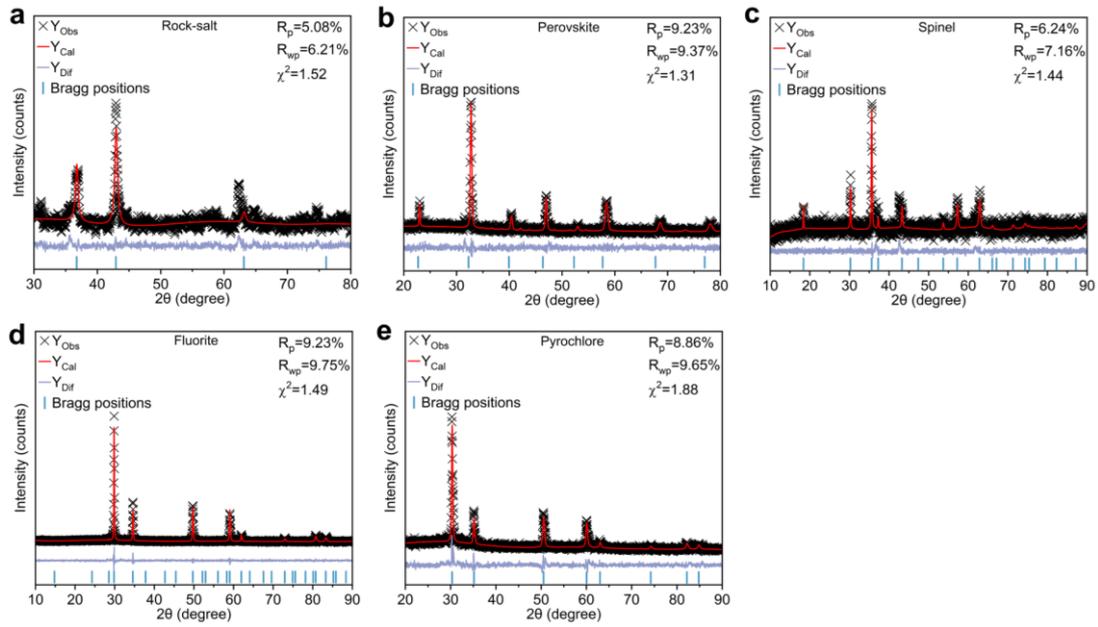

**Fig. S9** XRD Rietveld refinements of the as-synthesized HEO samples. (**a**) high-entropy rock-salt oxides. (**b**) high-entropy perovskite oxides. (**c**) high-entropy spinel oxides. (**d**) high-entropy fluorite oxides. (**e**) high-entropy pyrochlore oxides.

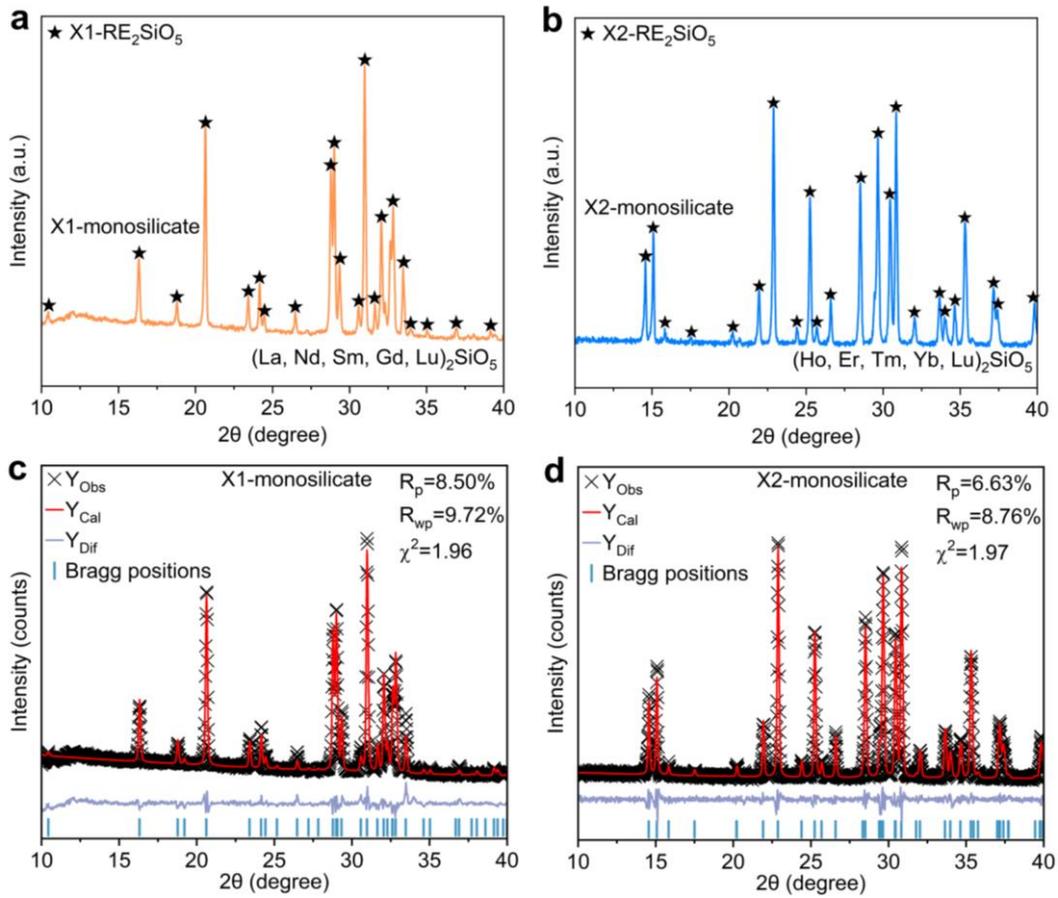

**Fig. S10** XRD patterns and corresponding Rietveld refinements of the as-synthesized X1- (**a, c**) and X2-type high-entropy RE monosilicate (**b, d**) samples.

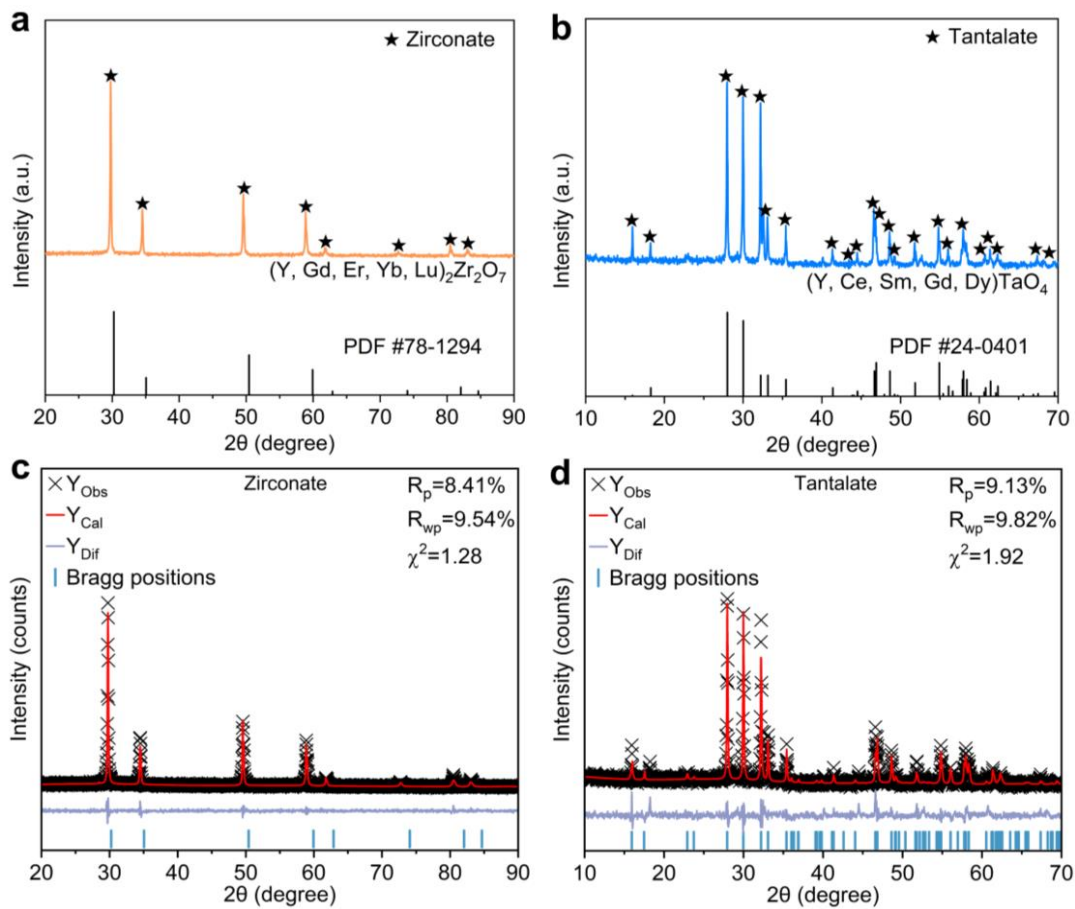

**Fig. S11** XRD patterns and corresponding Rietveld refinements of the as-synthesized high-entropy RE zirconate (**a, c**) and high-entropy RE tantalate (**b, d**) samples.

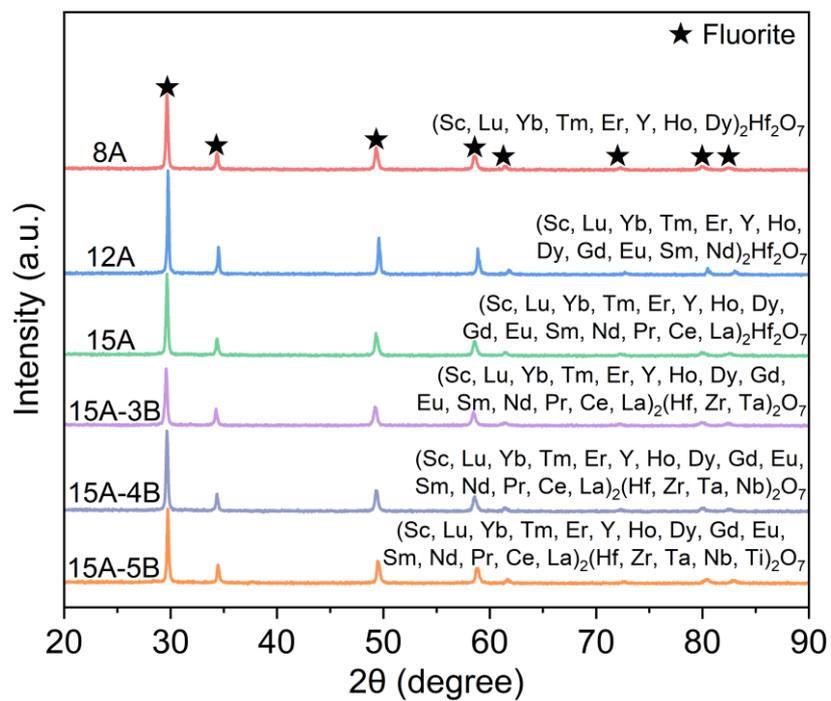

**Fig. S12** XRD patterns of the as-synthesized HEREH samples. A and B indicate that the cations occupy the A site and B site, respectively.

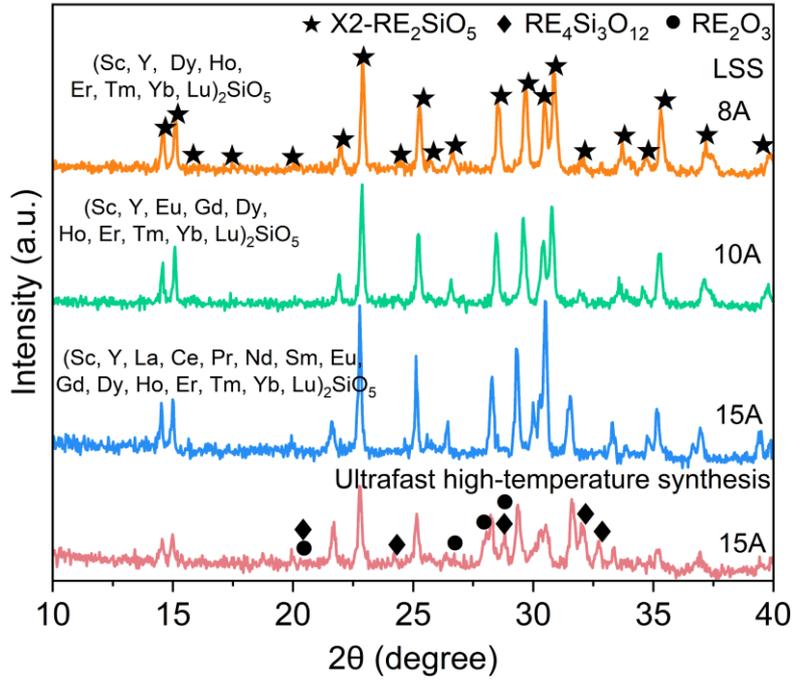

**Fig. S13** XRD patterns of the as-synthesized HERED samples through LSS and ultrafast high-temperature synthesis.

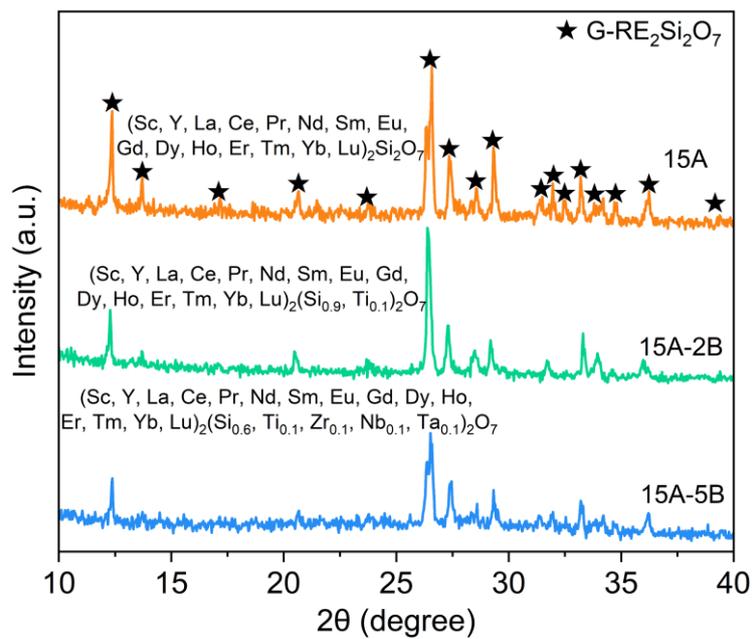

**Fig. S14** XRD patterns of the as-synthesized G-type HERED samples. A and B indicate that the cations occupy A site and B site, respectively.

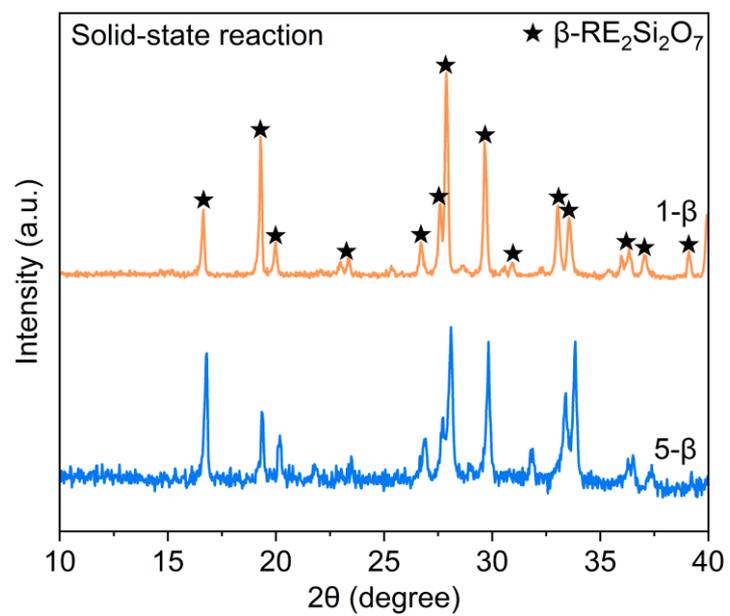

**Fig. S15** XRD patterns of the $Yb_2Si_2O_7$ and 5-cation HERED samples synthesized through conventional solid-state reaction.

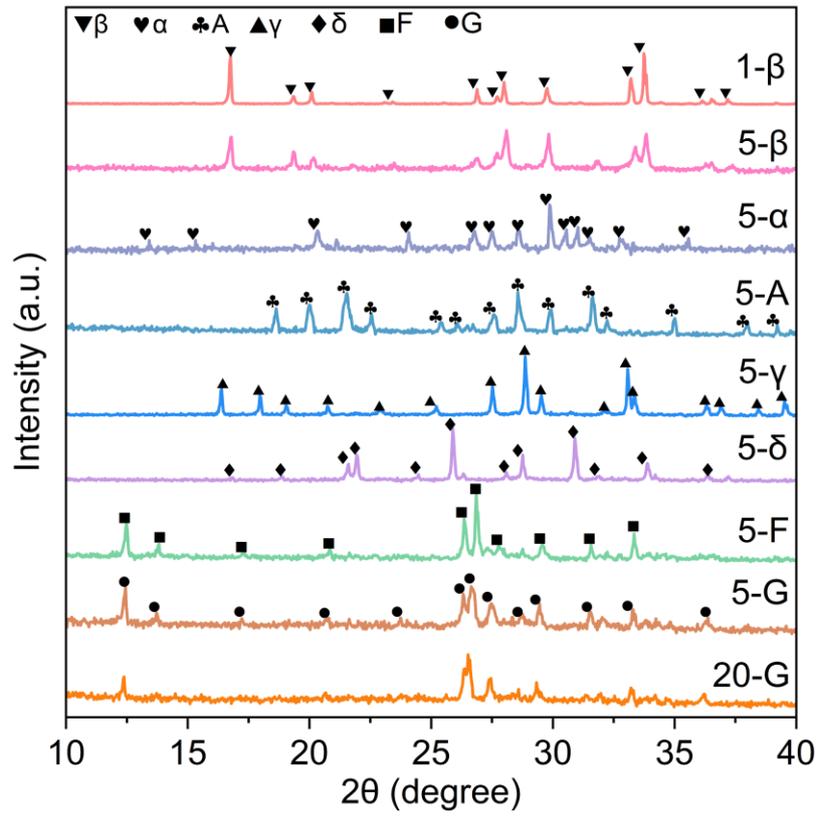

**Fig. S16** XRD patterns of the $Yb_2Si_2O_7$ and a series of HERED samples synthesized by LSS.

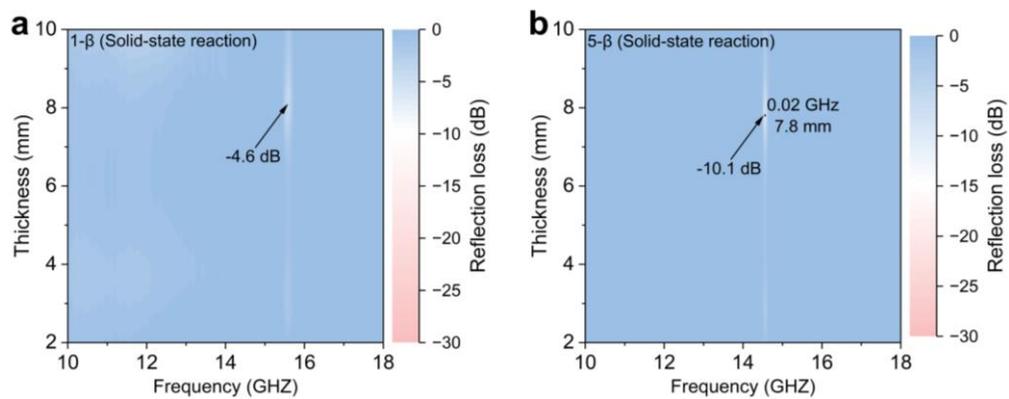

**Fig. S17** RL diagrams of the as-synthesized $Yb_2Si_2O_7$ (**a**) and 5-cation HERED (**b**) samples through the conventional solid-state reaction.

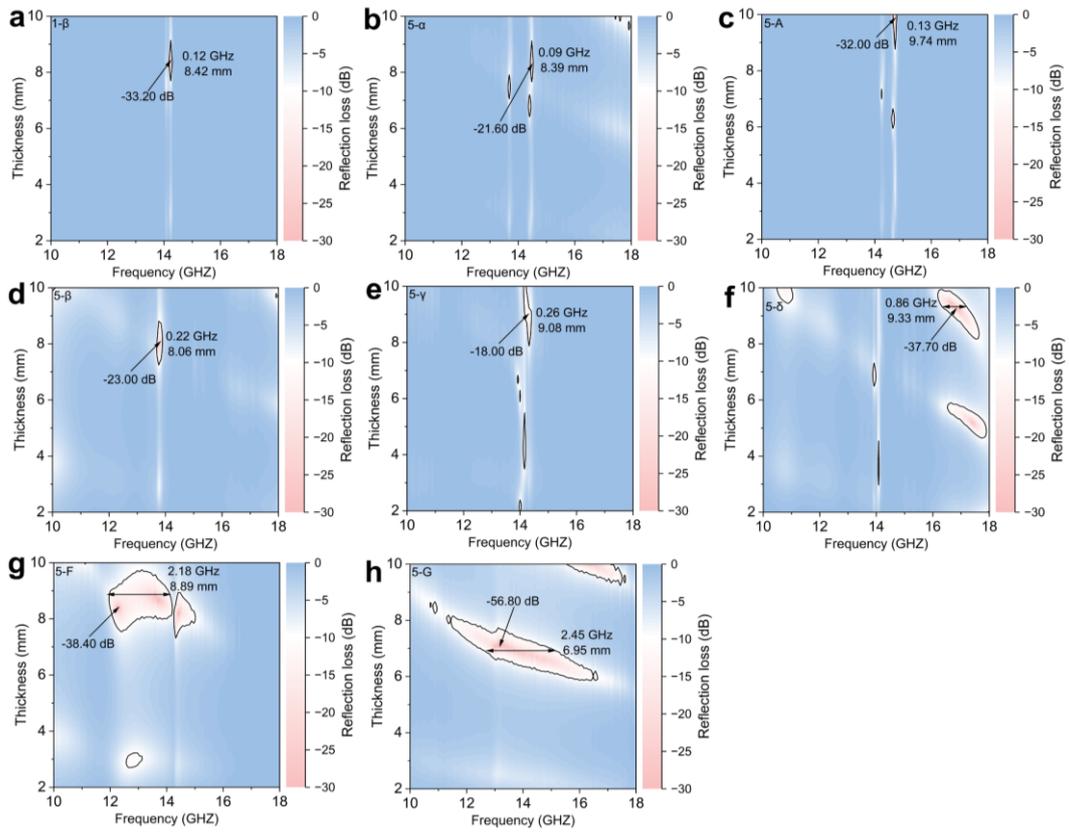

**Fig. S18** RL diagrams of the as-synthesized HERED samples through LSS. (**a**) 1-β. (**b**) 5-α. (**c**) 5-A. (**d**) 5-β. (**e**) 5-γ. (**f**) 5-δ. (**g**) 5-F. (**h**) 5-G.

Table S1 Refined lattice parameters of the as-synthesized HERED samples.

| Sample | Structure | Space group | $a$ (Å) | $b$ (Å) | $c$ (Å) | $\alpha$ (°) | $\beta$ (°) | $\gamma$ (°) |
|---|---|---|---|---|---|---|---|---|
| β-HERED-1 | Monoclinic | $C2/m$ | 6.84 | 8.92 | 4.73 | 90 | 101.87 | 90 |
| γ-HERED-2 | Monoclinic | $P2_1/c$ | 4.68 | 10.80 | 5.57 | 90 | 96.04 | 90 |
| δ-HERED-3 | Orthorhombic | $Pnma$ | 13.70 | 8.18 | 5.02 | 90 | 90 | 90 |
| F-HERED-4 | Triclinic | $P\bar{1}$ | 5.38 | 8.54 | 12.86 | 90.05 | 91.04 | 90.85 |
| G-HERED-5 | Monoclinic | $P2_1/n$ | 5.39 | 8.60 | 12.93 | 90 | 90.532 | 90 |
| A-HERED-6 | Tetragonal | $P4_1$ | 6.83 | 6.83 | 25.11 | 90 | 90 | 90 |
| α-HERED-7 | Triclinic | $P\bar{1}$ | 6.94 | 7.15 | 11.81 | 91.95 | 92.26 | 91.97 |

**Table 2** Atomic ratios of elements in the as-synthesized HERED samples.

| Sample | Element (at.%) | | | |
|---|---|---|---|---|
| β-HERED-1 | Sc | Eu | Gd | Yb |
| | 24.87 | 25.14 | 25.53 | 24.46 |
| γ-HERED-2 | Ce | Tm | Yb | Lu |
| | 25.21 | 25.39 | 24.77 | 24.63 |
| δ-HERED-3 | La | Tm | Yb | Lu |
| | 24.43 | 25.06 | 24.85 | 25.66 |
| G-HERED-5 | La | Ce | Pr | Lu |
| | 25.12 | 24.41 | 25.18 | 25.29 |

**Table S3** Refined lattice parameters of the as-synthesized HEO samples.

| Sample | Component | Structure | Space group | $a$ (Å) | $b$ (Å) | $c$ (Å) |
|---|---|---|---|---|---|---|
| Rock-salt | (Mg, Co, Ni, Al, Fe)O | Cubic | $Fm\bar{3}m$ | 4.08 | / | / |
| Perovskite | Sr(Cr, Mn, Fe, Co, Ni)O$_3$ | Orthorhombic | $pnma$ | 5.45 | 7.75 | 5.49 |
| Spinel | (Cu, Cr, Mn, Fe, Co, Ni)$_3$O$_4$ | Cubic | $Fm\bar{3}m$ | 8.36 | / | / |
| Fluorite | (Dy, Ho, Er, Tm, Lu)$_2$Hf$_2$O$_7$ | Cubic | $Fm\bar{3}m$ | 10.37 | / | / |
| Pyrochlore | Y$_2$(Zr, Ti, Hf, Al, Si)$_2$O$_7$ | Cubic | $Fm\bar{3}m$ | 5.11 | / | / |

**Table S4** Atomic ratios of elements in the as-synthesized HEO samples.

| Sample | Element (at.%) | | | | | | |
|---|---|---|---|---|---|---|---|
| Perovskite | Sr | Cr | Mn | Fe | Co | Ni | O |
| | 19.46 | 4.24 | 4.19 | 4.06 | 3.97 | 4.31 | 59.77 |
| Spinel | Cu | Cr | Mn | Fe | Co | Ni | O |
| | 7.56 | 7.29 | 6.98 | 7.12 | 7.43 | 7.32 | 56.30 |
| Fluorite | Dy | Ho | Er | Tm | Lu | Hf | O |
| | 4.06 | 3.76 | 3.94 | 3.57 | 3.62 | 18.67 | 62.38 |

Table S5 Atomic ratios of elements in the as-synthesized HEREH samples.

| Sample | Element (at.%) | | | | |
|---|---|---|---|---|---|
| HEREH | Sc | Y | La | Ce | Pr |
| | 3.27 | 3.19 | 3.35 | 3.44 | 3.36 |
| | Nd | Sm | Eu | Gd | Dy |
| | 3.22 | 3.14 | 3.55 | 3.26 | 3.65 |
| | Ho | Er | Tm | Yb | Lu |
| | 3.18 | 3.49 | 3.26 | 3.57 | 3.48 |
| | Ti | Zr | Nb | Ta | Hf |
| | 9.78 | 9.51 | 9.93 | 10.15 | 10.22 |

Table S6 Atomic ratios of elements in the as-synthesized HERED samples.

| Sample | Element (at.%) | | | | |
|---|---|---|---|---|---|
| HERED | Sc | Y | La | Ce | Pr |
| | 3.34 | 3.28 | 3.59 | 3.46 | 3.28 |
| | Nd | Sm | Eu | Gd | Dy |
| | 3.19 | 3.31 | 3.29 | 3.54 | 3.27 |
| | Ho | Er | Tm | Yb | Lu |
| | 3.15 | 3.46 | 3.32 | 3.44 | 3.27 |
| | Ti | Zr | Nb | Ta | Si |
| | 4.57 | 4.82 | 5.31 | 5.19 | 29.92 |